\documentclass[11pt, a4paper]{article}
\pdfoutput=1
\usepackage{jcappub}

\usepackage{booktabs}
\usepackage{multirow}
\usepackage{amssymb}
\usepackage[export]{adjustbox}
\usepackage{floatrow}
\newfloatcommand{capbtabbox}{table}[][3.5in]
\newfloatcommand{capbfigbox}{figure}[][2.3in]
\usepackage{blindtext}
\usepackage{amsmath}
\usepackage{booktabs}
\usepackage{aas_macros}

\def\lsim{\mathrel{\raise.3ex\hbox{$<$\kern-.75em\lower1ex\hbox{$\sim$}}}}
\def\gsim{\mathrel{\raise.3ex\hbox{$>$\kern-.75em\lower1ex\hbox{$\sim$}}}}

\usepackage{xcolor}
\usepackage[normalem]{ulem}
\usepackage{soul}

\begin{document}

\hspace*{101.3mm}{\large \tt IFIC/16-66}\\
\hspace*{110mm}{\large \tt FERMILAB-PUB-16-370-A}

\vskip 0.2in

\title{Toward (Finally!) Ruling Out Z and Higgs Mediated Dark Matter Models} 

\author[a,b]{Miguel Escudero,}\note{ORCID: http://orcid.org/0000-0002-4487-8742}
\emailAdd{miguel.escudero@ific.uv.es}
\author[c]{Asher Berlin,}\note{ORCID: http://orcid.org/0000-0002-1156-1482}
\emailAdd{berlin@uchicago.edu}
\author[b,d,e]{Dan Hooper}\note{ORCID: http://orcid.org/0000-0001-8837-4127}
\emailAdd{dhooper@fnal.gov}
\author[d]{and Meng-Xiang Lin}\note{ORCID: http://orcid.org/0000-0003-2908-4597}
\emailAdd{mxlin@uchicago.edu}

\affiliation[a]{Instituto de F\'{\i}sica Corpuscular (IFIC)$,$ CSIC-Universitat de Val\`encia$,$ Apartado de Correos 22085$,$ E-46071 Valencia$,$ Spain}
\affiliation[b]{Fermi National Accelerator Laboratory, Center for Particle
Astrophysics, Batavia, IL 60510}
\affiliation[c]{University of Chicago, Department of Physics, Chicago, IL 60637}
\affiliation[d]{University of Chicago, Department of Astronomy and Astrophysics, Chicago, IL 60637}
\affiliation[e]{University of Chicago, Kavli Institute for Cosmological Physics, Chicago, IL 60637}

\abstract{In recent years, direct detection, indirect detection, and collider experiments have placed increasingly stringent constraints on particle dark matter, exploring much of the parameter space associated with the WIMP paradigm. In this paper, we focus on the subset of WIMP models in which the dark matter annihilates in the early universe through couplings to either the Standard Model $Z$ or the Standard Model Higgs boson. Considering fermionic, scalar, and vector dark matter candidates within a model-independent context, we find that the overwhelming majority of these dark matter candidates are already ruled out by existing experiments. In the case of $Z$ mediated dark matter, the only scenarios that are not currently excluded are those in which the dark matter is a fermion with an axial coupling and with a mass either within a few GeV of the $Z$ resonance ($m_{\rm DM} \simeq m_Z/2$) or greater than 200 GeV, or with a vector coupling and with $m_{\rm DM} > 6$ TeV. Several Higgs mediated scenarios are currently viable if the mass of the dark matter is near the Higgs pole ($m_{\rm DM} \simeq m_H/2$). Otherwise, the only scenarios that are not excluded are those in which the dark matter is a scalar (vector) heavier than 400 GeV (1160 GeV) with a Higgs portal coupling, or a fermion with a pseudoscalar (CP violating) coupling to the Standard Model Higgs boson. With the exception of dark matter with a purely pseudoscalar coupling to the Higgs, it is anticipated that planned direct detection experiments will probe nearly the entire range of models considered in this study. 
}

\maketitle

\section{Introduction}

Over the past few decades, the WIMP paradigm has dominated the theoretical and experimental landscape of dark matter. Interest in dark matter in the form of weakly interacting massive particles (WIMPs) has been motivated in large part by the realization that a generic stable particle with an electroweak scale mass and interactions will freeze-out in the early universe with a thermal relic abundance that is comparable to the measured cosmological dark matter density. And although there are many electroweak processes through which a WIMP could potentially annihilate, none are as ubiquitous across the landscape of dark matter models than those which result from couplings between the dark matter  and the Standard Model (SM) $Z$ or Higgs bosons. 

As direct detection, indirect detection, and collider searches for dark matter have progressed, the WIMP paradigm has become increasingly well explored and constrained. And although there remain many viable WIMP models, important experimental benchmarks have been reached, providing us with valuable information pertaining to the identity of our universe's dark matter. In this paper, we focus on the subset of models in which the dark matter annihilates through the exchange of the $Z$ or the Higgs boson. In scenarios outside of this subset of models, WIMPs must annihilate through the exchange of particles beyond the SM if they are to avoid being overproduced in the early universe. 

The remainder of this article is structured as follows. In Sec.~\ref{Z}, we consider dark matter that is mediated by $Z$ exchange, discussing fermionic, scalar and vector dark matter candidates. We find that a very significant part of this parameter space is ruled out by a combination of constraints from direct detection experiments (LUX, PandaX-II) and measurements from LEP of the invisible width of the $Z$. The only scenarios which remain viable at this time are those with a fermionic dark matter candidate with a nearly pure axial coupling to the $Z$ and with a mass that lies within either a few GeV of the $Z$ pole ($m_{\rm DM} \simeq 40-48$ GeV) or that is heavier than 200 GeV, or a fermion with a vector coupling to the $Z$ and that is heavier than 6 TeV. Much of this parameter space is expected to be tested in the near future with direct detection experiments such as XENON1T. In Sec.~\ref{H}, we consider fermionic, scalar and vector dark matter candidates that are coupled to the SM Higgs boson. Across this class of models, we again find that the overwhelming majority of the parameter space is experimentally excluded, with the exception of scenarios in which the dark matter lies near the Higgs pole ($m_{\rm DM} \simeq m_H/2$), the dark matter is a scalar (vector) heavier than 400 GeV (1160 GeV) with a Higgs portal coupling, or the dark matter is a fermion with largely pseudoscalar couplings to the SM Higgs boson. In Sec.~\ref{caveats} we discuss some caveats to our conclusions, including scenarios with a non-standard cosmological history, or models in which the dark matter coannihilates with another particle species in the early universe. We summarize our results and conclusions in Sec.~\ref{conclusions}.

\section{$Z$ Mediated Dark Matter}
\label{Z}

\subsection{Fermionic dark matter}

\begin{figure*}
\begin{center}
\begin{tabular}{cc}
    \includegraphics[width=0.48\textwidth]{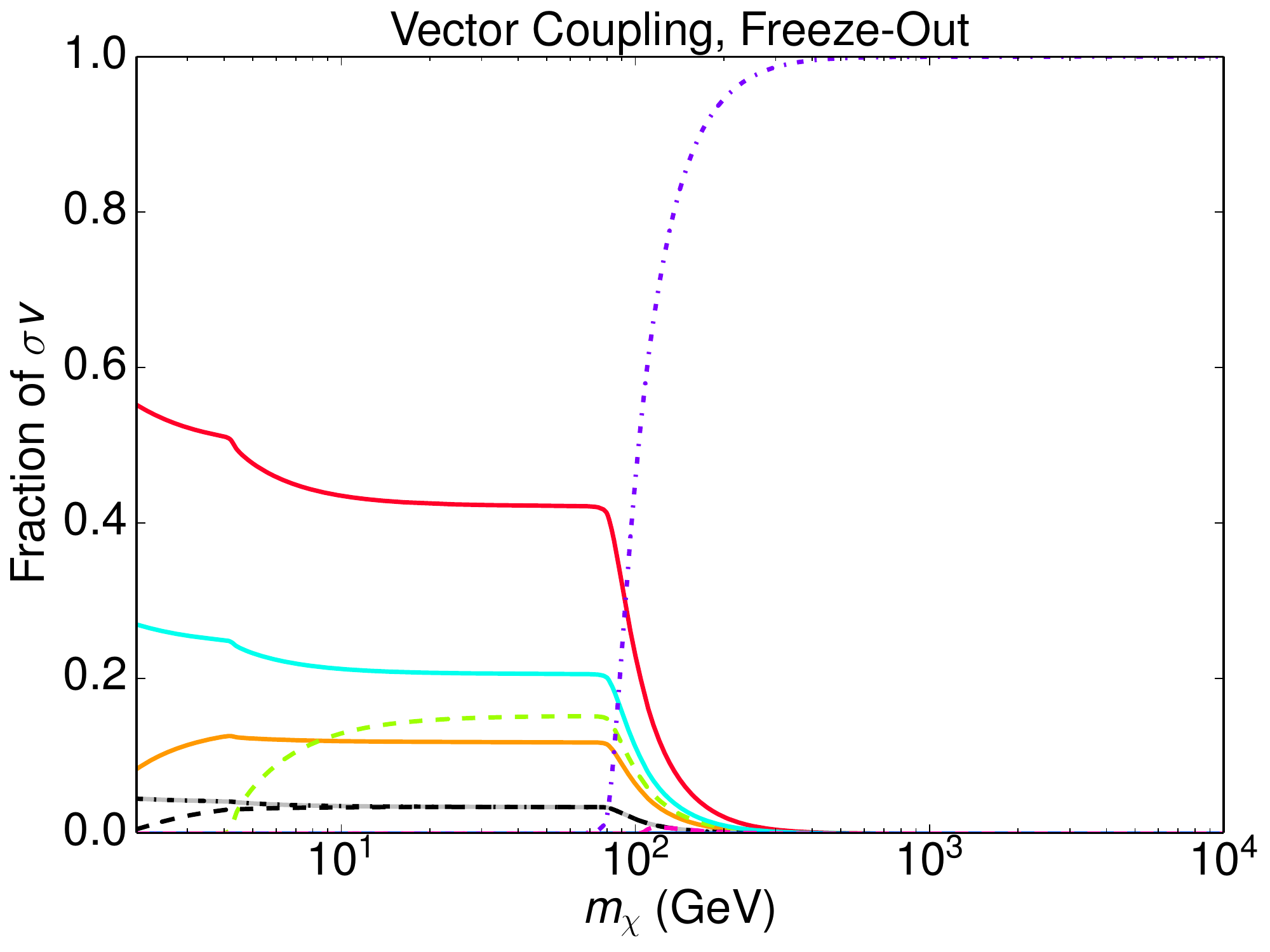} &   \includegraphics[width=0.48\textwidth]{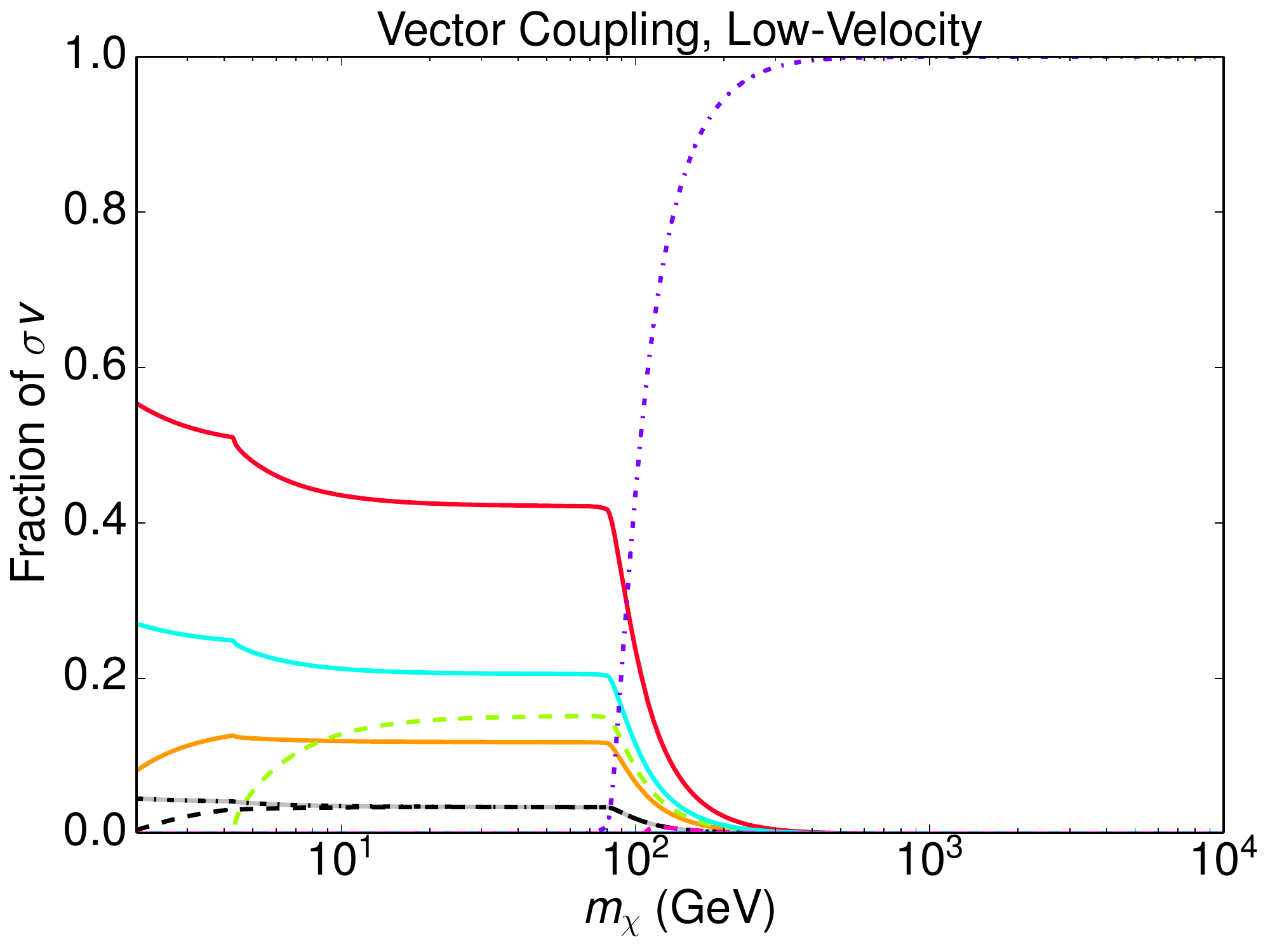} \\
    \includegraphics[width=0.48\textwidth]{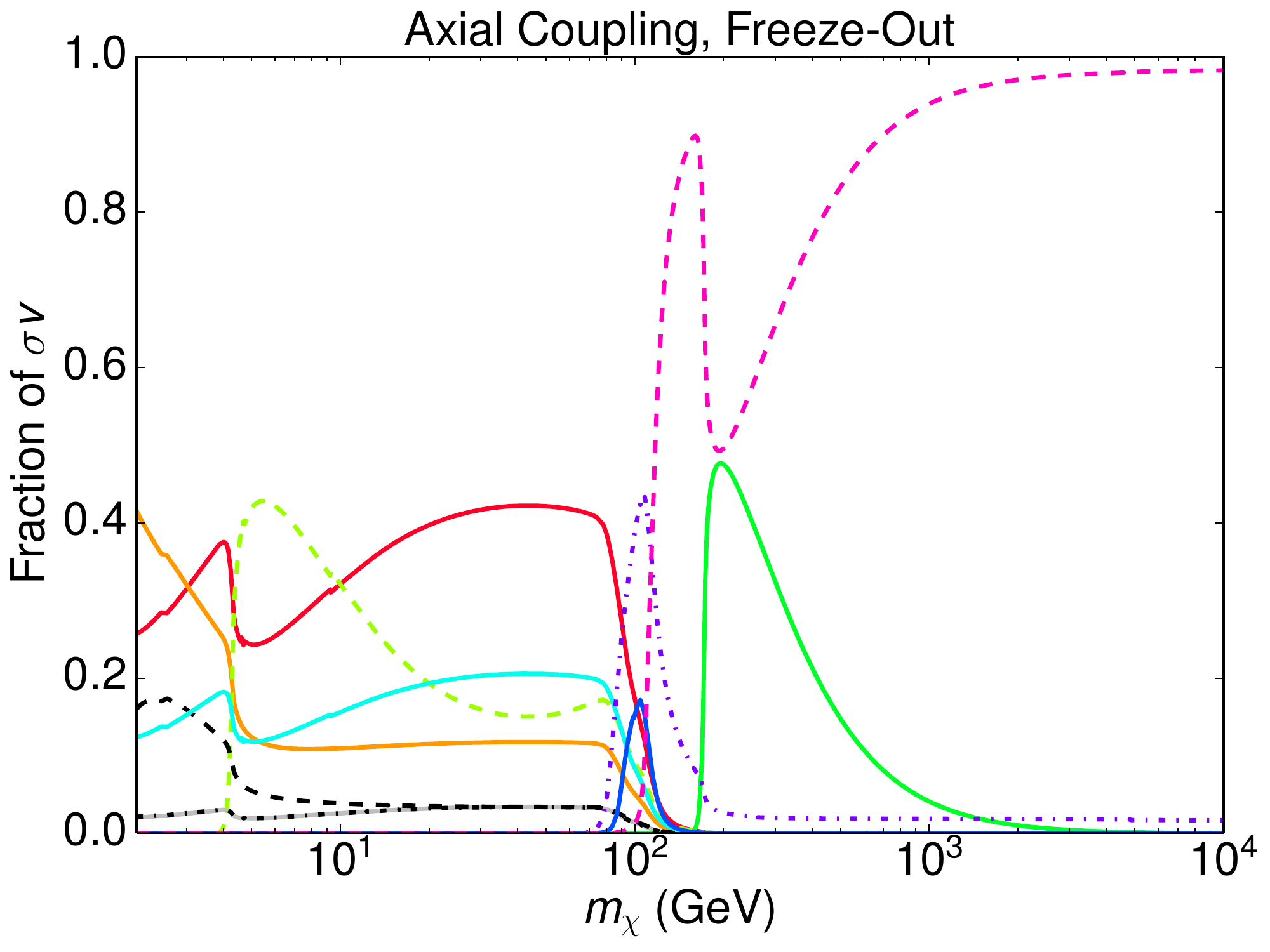} &    \includegraphics[width=0.48\textwidth]{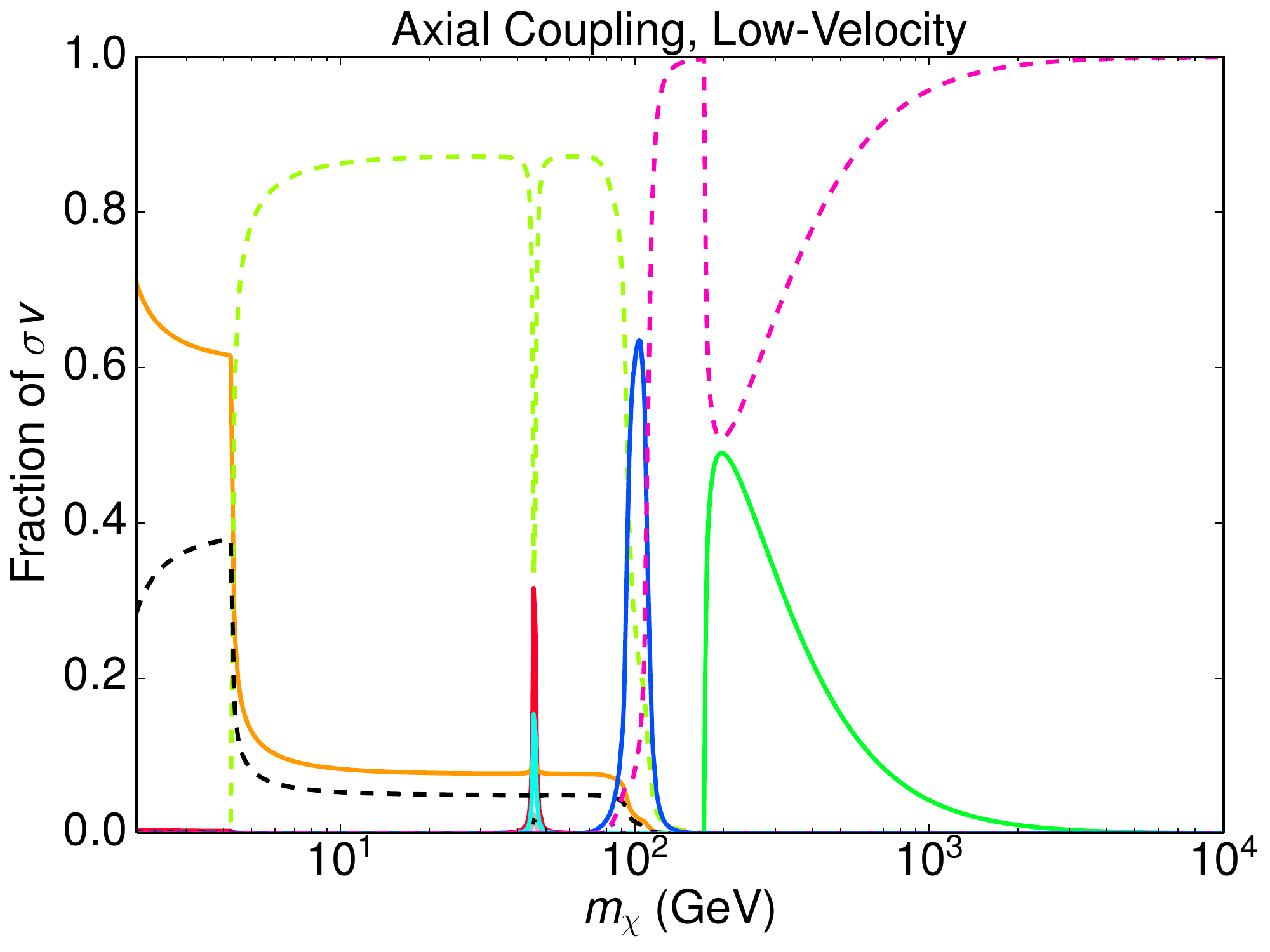}
\end{tabular}
\includegraphics[width=0.98\textwidth]{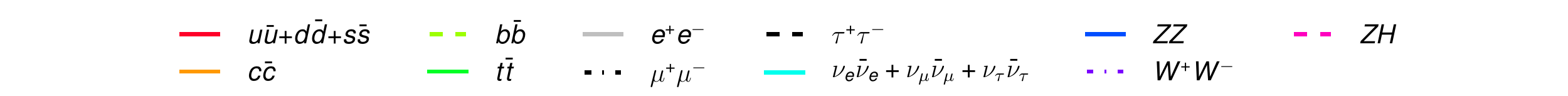} 
\caption[]{The fraction of dark matter annihilations that proceed to each final state, as evaluated at the temperature of thermal freeze-out (left) and at $v=10^{-3}\,c$, as is typically relevant for indirect detection (right). The upper and lower frames correspond to dark matter in the form of a fermion with purely vector or purely axial couplings to the $Z$, respectively.}
\label{fig:contributions}
\end{center}
\end{figure*}

We begin by considering a dark matter candidate, $\chi$, which is either a Dirac or a Majorana fermion with the following interactions with the SM $Z$:
\begin{equation}\label{eq:Zfermion}
\mathcal{L} \supset \left[ a \bar{\chi} \gamma^\mu ( g_{\chi  v } + g_{\chi a} \gamma^5 ) \chi  \right] Z_\mu, \\
\end{equation}
where $a=1 \,(1/2)$ in the Dirac (Majorana) case, and $g_{\chi  v }$ and $g_{\chi  a}$ are the vector and axial couplings of the dark matter, respectively. Note that $g_{\chi v}$ is necessarily equal to zero in the Majorana case. These couplings allow the dark matter to annihilate through the $s$-channel exchange of the $Z$, into pairs of SM fermions or, if the dark matter is heavy enough, into $ZZ$, $W^+ W^-$ or $Zh$ final states. In Fig.~\ref{fig:contributions} we plot the fraction of annihilations which proceed to each final state, as evaluated in the early universe (at the temperature of thermal freeze-out) and for $v=10^{-3} \,c$ (as is typically relevant for indirect detection). Throughout this paper, unless otherwise stated, we use version 4.2.5 of the publicly available code MicrOMEGAS~\cite{Belanger:2014vza} to calculate all annihilation cross sections, thermal relic abundances, and elastic scattering cross sections.

\begin{figure*}
\begin{center}
    \includegraphics[width=0.49\textwidth]{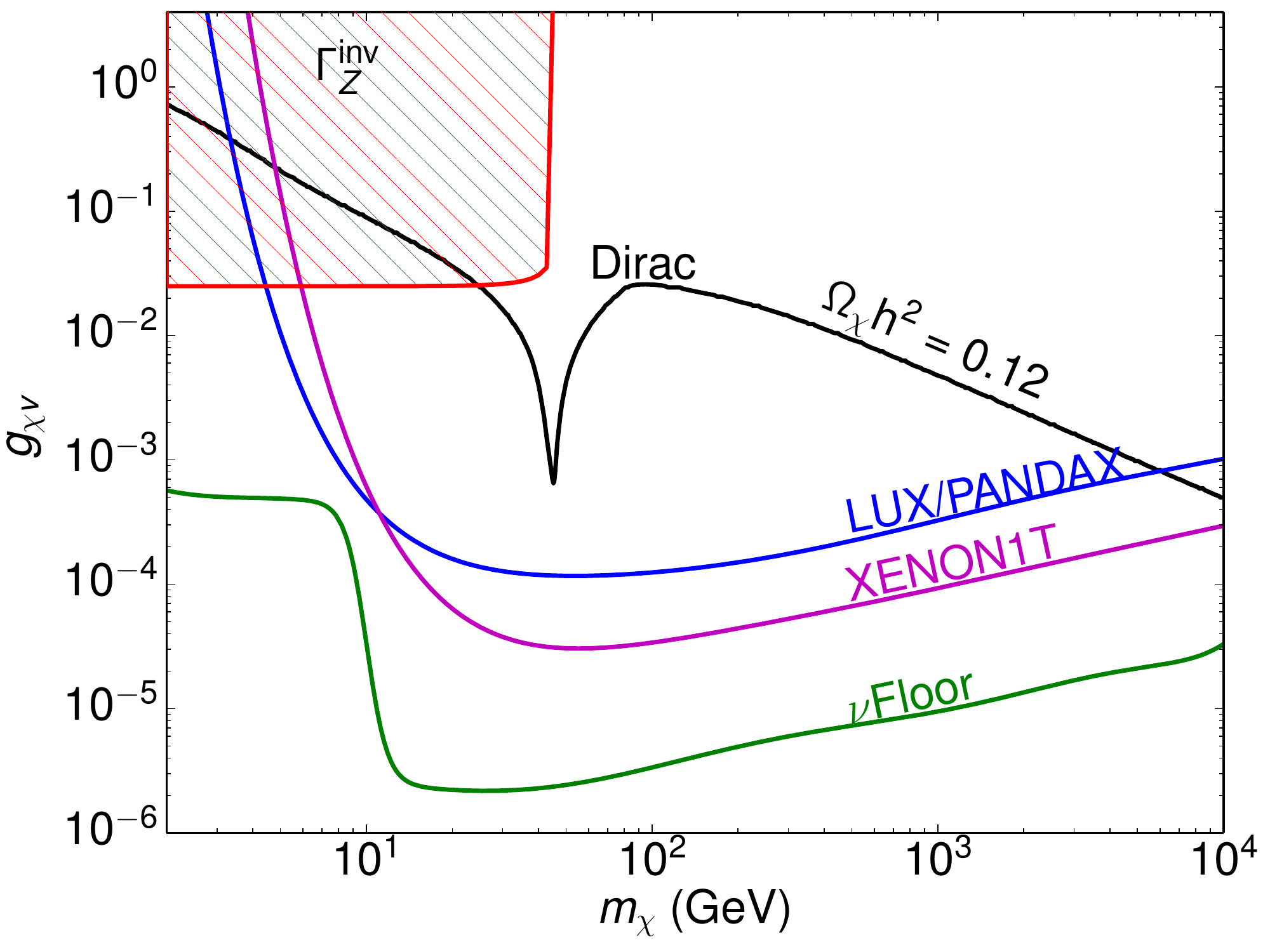}
    \includegraphics[width=0.49\textwidth]{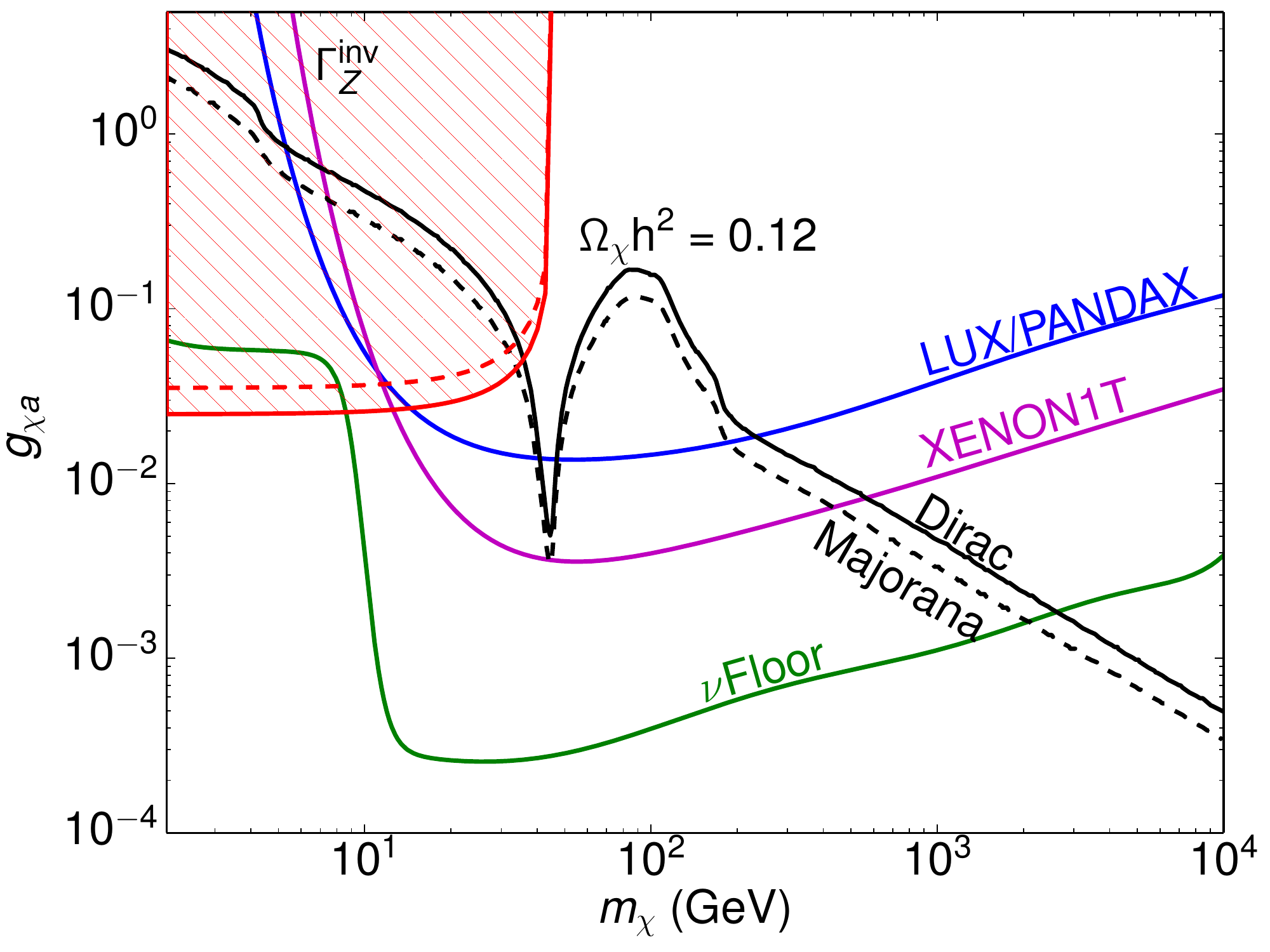}
\caption[]{Constraints on the mass and couplings of a fermionic dark matter candidate that annihilates through the $Z$. The solid black contours indicate the value of the coupling for which the thermal relic abundance matches the measured cosmological dark matter density, $\Omega_{\chi} h^2 = 0.12$. The shaded regions are excluded by the measurement of the invisible $Z$ width. The left and right frames depict the cases of a purely vector or axial coupling between the dark matter and the $Z$, respectively. The vast majority of this parameter space is excluded by the current constraints from LUX and PandaX-II~\cite{Akerib:2016vxi,Tan:2016zwf}, and much of the currently viable parameter space is expected to be probed in the near future by XENON1T~\cite{Aprile:2015uzo}.}
\label{fig:Constraints}
\end{center}
\end{figure*}

In Fig.~\ref{fig:Constraints}, we explore and summarize the parameter space within this class of models. In each frame, the solid black line represents the value of the dark matter's coupling to the $Z$ ($g_{\chi v}$ or $g_{\chi a}$) for which the calculated thermal relic abundance is equal to the measured cosmological dark matter density, $\Omega_{\chi} h^2 = 0.1198 \pm 0.0015$~\cite{Ade:2015xua}. If $m_\chi < m_Z/2$, we can further restrict the couplings of the dark matter using the measurement of the invisible $Z$ width. The predicted contribution from $Z$ decays to dark matter is in this case is given by:
\begin{eqnarray}
\Gamma(Z\to\chi\bar{\chi}) = \frac{a \, m_Z}{12 \pi} \left(1-\frac{4m_\chi^2}{m_Z^2}\right)^{1/2} \left[g_{\chi a}^2\left(1-\frac{4m_\chi^2}{m_Z^2}\right)+g_{\chi v}^2\left(1+\frac{2m_\chi^2}{m_Z^2}\right)\right],
\end{eqnarray}
where again $a=1 (1/2)$ for dark matter that is a Dirac (Majorana) fermion. In the shaded regions appearing in the upper left corner of each frame of Fig.~\ref{fig:Constraints}, the predicted invisible width of the $Z$ exceeds the value measured at LEP by more the $2\sigma$, corresponding to a contribution of $\Gamma_Z^{\rm inv} > 1.5$ MeV~\cite{Agashe:2014kda}. Combined with relic abundance considerations, this constraint translates to $m_{\chi} > 25$ GeV (32 GeV) for the case of a purely vector (axial) coupling to the $Z$.

Direct detection experiments provide a powerful test of dark matter candidates with non-negligible couplings to the $Z$. After integrating out the $Z$, the effective interaction relevant for dark matter scattering with nuclei is given by: 
\begin{eqnarray}
\mathcal{L} = \frac{1}{m_Z^2} \left[\bar{\chi}\gamma^\mu (g_{\chi v} + g_{\chi a} \gamma^5)\chi \right]\left[\bar{q}\gamma^\mu (g_{q v} + g_{q a} \gamma^5)q \right],
\end{eqnarray}
where $g_{u v}=g_2 (\frac{1}{4 c_W}-\frac{2 s^2_W}{3 c_W})$, $g_{d v}=g_2 (-\frac{1}{4 c_W}+\frac{s^2_W}{3 c_W})$, $g_{u a}=-g_2/4 c_W$, $g_{d a}=g_2/4 c_W$, etc. are the couplings of the $Z$ to Standard Model quarks.

At low energies, $\bar{\Psi}\gamma^i\Psi \rightarrow 0$ and $\bar{\Psi}\gamma^0 \gamma^5\Psi \rightarrow 0$, and thus only vector-vector and axial-axial interactions are not suppressed by powers of velocity or momentum transfer. These interactions lead to spin-independent and spin-dependent scattering cross sections, respectively. The solid blue curves shown in Fig.~\ref{fig:Constraints} represent the current limits on the dark matter's coupling to the $Z$, as derived from the results of the direct detection experiment LUX~\cite{Akerib:2016vxi} (the PandaX-II experiment has placed a constraint that is only slightly weaker~\cite{Tan:2016zwf}).\footnote{Although the spin-independent constraints from direct detection experiments are generally presented for the case of equal couplings to protons and neutrons, we have translated these results to apply to the models at hand. It is interesting to note that a cancellation in the vector couplings of the $Z$ to up and down quarks leads to a suppression in the effective coupling to protons. In particular, $Z$ exchange leads to the following ratio of cross sections with neutrons and protons: $\sigma_n/\sigma_p \approx (2 g_{dv} + g_{uv})^2/(2g_{uv}+g_{dv})^2 \approx 180$. We also note that since xenon contains isotopes with an odd number of neutrons (${}^{129}$Xe and ${}^{131}$Xe with abundances of 29.5\% and 23.7\%, respectively), this target is quite sensitive to spin-dependent WIMP-neutron scattering. To constrain spin-dependent scattering, we converted the results of the most recent spin-independent analysis presented by the LUX collaboration~\cite{Akerib:2016vxi}.}

Together, these constraints rule out the majority of the parameter space for fermionic dark matter candidates that annihilate through $Z$ exchange. After accounting for these constraints, we find that an acceptable thermal relic abundance can be obtained only in the near-resonance case~\cite{Matsumoto:2016hbs,Matsumoto:2014rxa,Arcadi:2014lta} ($m_\chi = m_Z/2$) or for $m_\chi \gsim 200$ GeV with $g_{\chi a} \gg g_{\chi v}$, or for $m_\chi \gsim 6$ TeV. Furthermore, with the exception of $m_\chi \gsim 500$ GeV with $g_{\chi a} \gg g_{\chi v}$, we expect that the remaining parameter space will be probed in the near future by direct detection experiments such as XENON1T~\cite{Aprile:2015uzo}. We point out that for fermionic dark matter heavier than several TeV, perturbative unitarity is lost, and higher dimension operators such as those ones considered in Ref.~\cite{deSimone:2014pda} may become relevant for the phenomenology. It is interesting to note that within the context of the MSSM, a bino-like neutralino (with a subdominant higgsino fraction) can possess the characteristics found within this scenario~\cite{Hooper:2013qjx}.

\begin{figure}[t]
\begin{center}
\begin{tabular}{cc}
\includegraphics[width=0.49\textwidth]{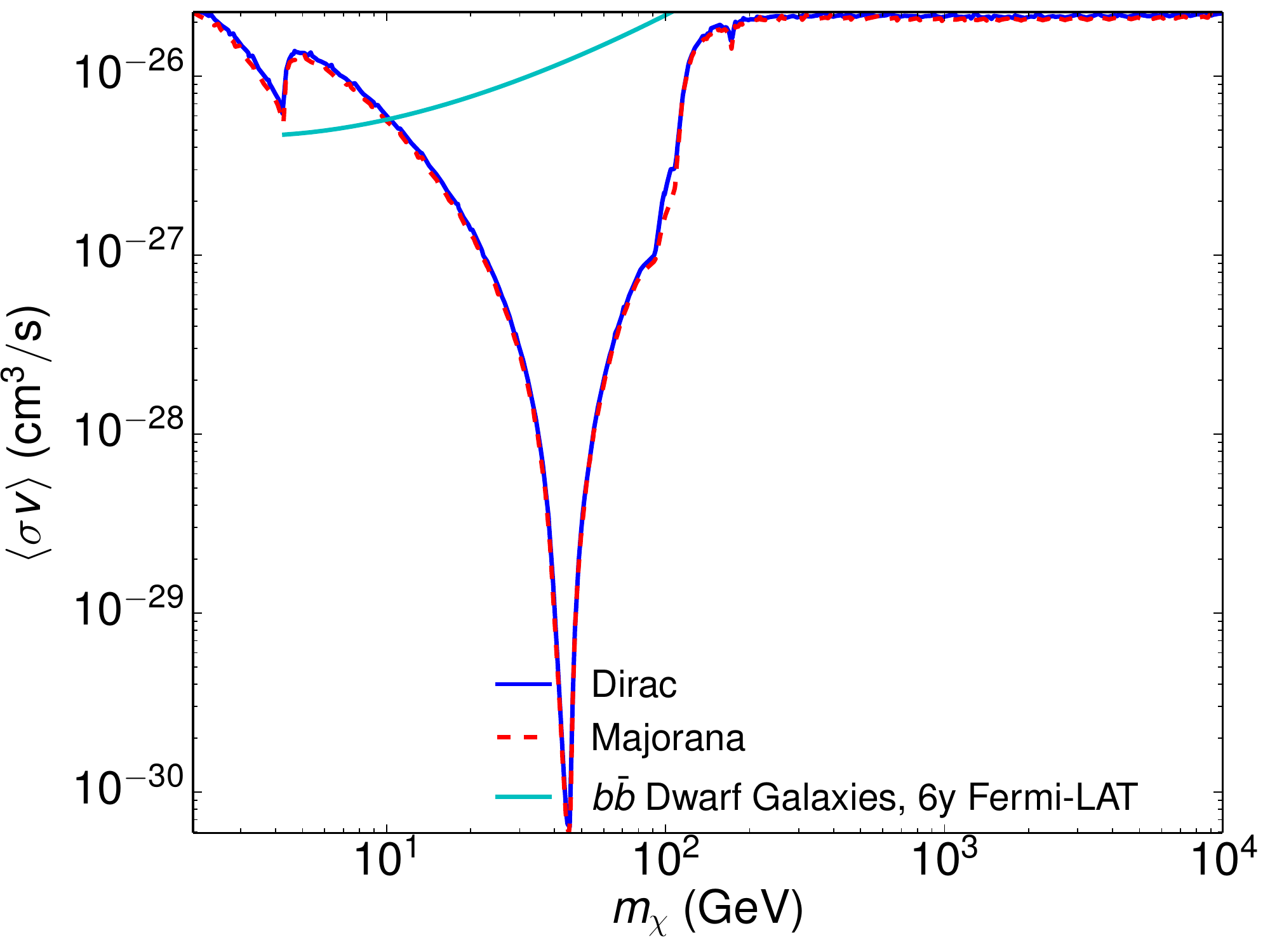}
\end{tabular}
\caption[]{The effective low-velocity annihilation cross section (relevant for indirect detection) for Dirac or Majorana dark matter with an axial coupling to the $Z$. We note that for the masses in the range not yet excluded by LUX or PandaX-II, this cross section is well below the constraints derived from Fermi and other existing indirect detection experiments~\cite{Hooper:2012sr,Ackermann:2015zua,Bergstrom:2013jra,Giesen:2015ufa,Cirelli:2013hv}. We also show the current constraint from Fermi's observation of dwarf spheroidal galaxies~\cite{Ackermann:2015zua}.}
\label{fig:Majorana}
\end{center}
\end{figure}

In the narrow region of viable parameter space found near the $Z$ pole, the dark matter in this class of models annihilates with a cross section that is chirality suppressed in the low-velocity limit, $\sigma v \propto (m_f/m_{\chi})^2$, leading such annihilations to proceed mostly to $b\bar{b}$ final states. In this mass range, the low-velocity cross section is sensitive to the value of the dark matter's mass, but consistently below the reach of planned indirect detection experiments (for analytic expressions of this cross section, see the Appendix of Ref.~\cite{Berlin:2014tja}). In Fig.~\ref{fig:Majorana}, we plot the effective low-velocity annihilation cross section (as relevant for indirect detection) for fermionic dark matter with an axial coupling to the $Z$.\footnote{By ``effective'' annihilation cross section we denote the value for the case of identical annihilating particles (Majorana fermions). For a Dirac fermion (or a complex boson), the actual particle-antiparticle annihilation cross section is equal to twice this value.} These cross sections are well below the constraints derived from Fermi and other existing indirect detection experiments~\cite{Hooper:2012sr,Ackermann:2015zua,Bergstrom:2013jra,Giesen:2015ufa,Cirelli:2013hv}.

\subsection{Scalar dark matter}
\begin{figure}[t]
\begin{center}
\includegraphics[width=0.49\textwidth]{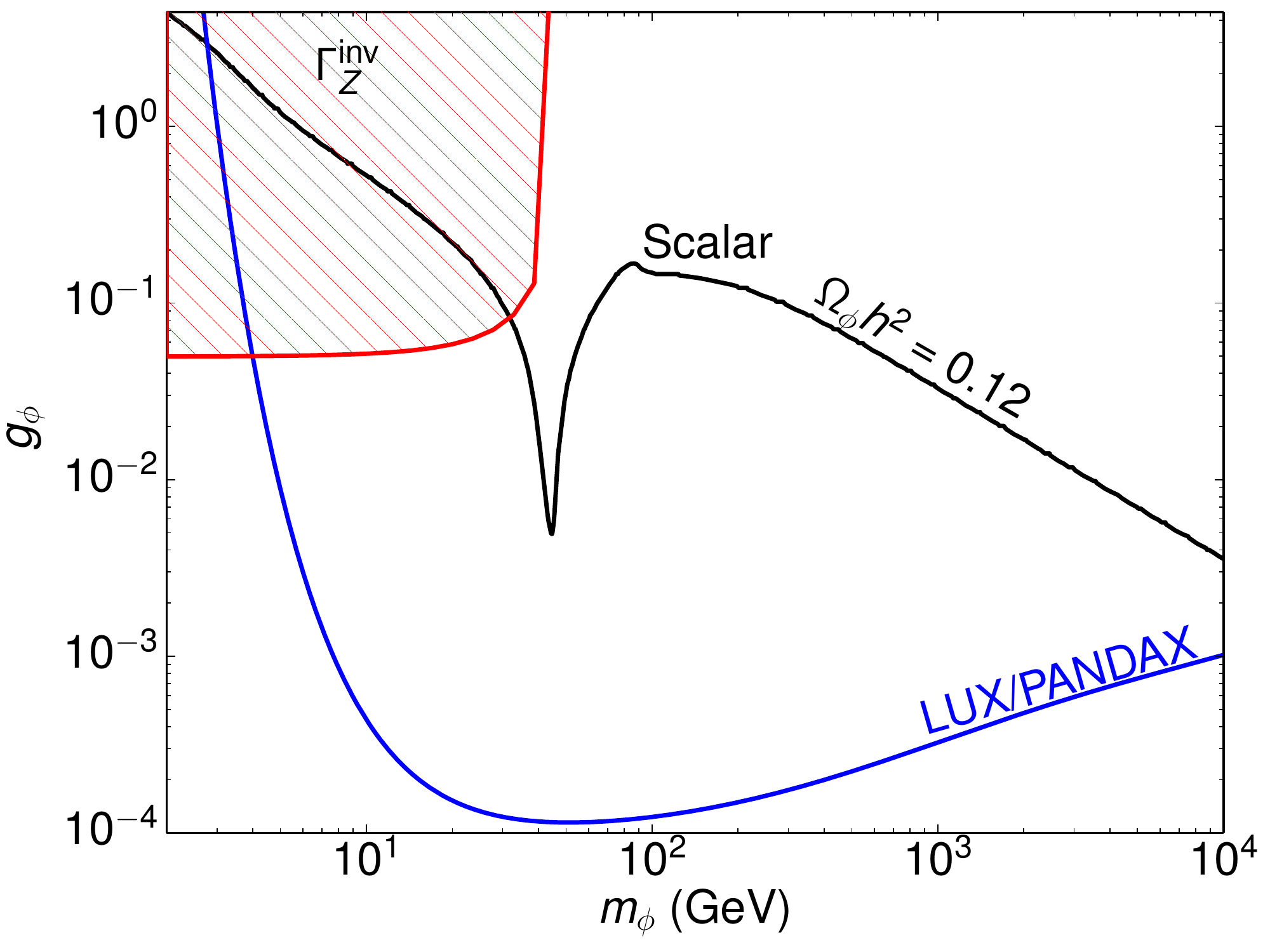}
\includegraphics[width=0.49\textwidth]{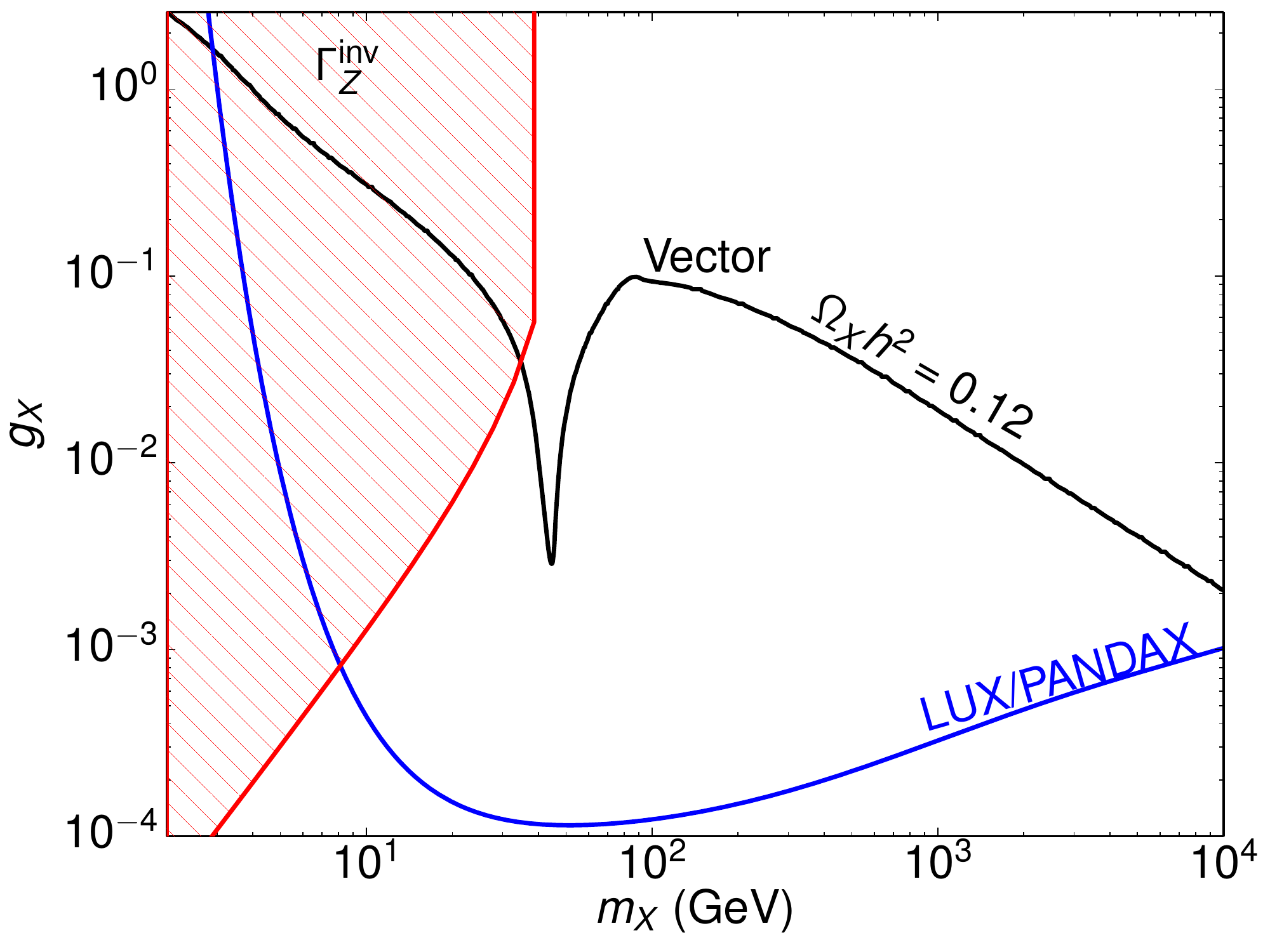}
\caption[]{Constraints on mass and couplings of a complex scalar (left frame) or complex vector (right frame) dark matter candidate which annihilates through the $Z$. The solid black contours indicate the value of the coupling for which the thermal relic abundance matches the measured cosmological dark matter density, $\Omega_{\chi} h^2 = 0.12$. The shaded regions are excluded by measurements of the invisible $Z$ width, and the regions above the solid blue line are excluded by the current constraints from LUX and PandaX-II~\cite{Akerib:2016vxi,Tan:2016zwf}. The entire parameter space in each of these scenarios is strongly ruled out.}
\label{fig:CS}
\end{center}
\end{figure}

%%%%%%%%%%%%%%%%%%%%%%%%%%%%%%%%%%%%%%%%%%%%%%%%%%%%%%
A complex scalar dark matter candidate, $\phi$, can couple to the $Z$ through the following interaction:  
\begin{equation}
\mathcal{L} \supset i\, g_\phi  \phi^\dagger  \overset{\leftrightarrow}{\partial_{\mu}}  \phi   Z^{\mu} + g^2_{\phi} \phi^2 Z^{\mu} Z_{\mu}. 
\end{equation}
The annihilation cross section to fermion pairs in this case is suppressed by two powers of velocity, and values of $g_{\phi}$ that lead to an acceptable thermal relic abundance are shown as a black solid line in the left frame of Fig.~\ref{fig:CS}. We also show in this figure the region of parameter space that is excluded by the measurement of the invisible width of the $Z$, which receives the following contribution in this case:
\begin{equation}
\Gamma (Z\to \phi\phi^\dagger)  = \frac{g_\phi^2 m_Z}{48 \pi }\left(1-\frac{4 m_\phi^2}{m_Z^2}\right)^{3/2}.
\end{equation}

In this model, there is an unsuppressed cross section for spin-independent elastic scattering with nuclei, leading to very stringent constraints from LUX and PandaX-II. In the left frame of Fig.~\ref{fig:CS}, we see that the entire parameter space in this scenario is strongly ruled out by a combination of constraints from LUX/PandaX-II and the invisible width of the $Z$.

\subsection{Vector dark matter}
%%%%%%%%%%%%%%

An interaction between the $Z$ and a spin-one dark matter candidate, $X$, can arise at tree-level only through a kinetic term. In this case, gauge invariance requires the interaction to take the following form:
\begin{equation}
\mathcal{L} \supset i \, g_X \, \left( Z^\mu \, X^{\nu \, \dagger} \partial_{\, [ \mu} X_{\nu \, ]} + X_\mu^\dagger X_\nu \, \partial^\mu Z^\nu \right) + \text{h.c.}
\end{equation}

In the right frame of Fig.~\ref{fig:CS}, we summarize the parameter space in this model. In particular, we apply constraints from the invisible width of the $Z$, which receives the following contribution in this case:
\begin{equation}
\Gamma (Z\to X X^\dagger)  = \frac{g_X^2 m_Z  \left(
1-8 r_{XZ}^2+28 r_{XZ}^4-48 r_{XZ}^6\right)\,(1-4r^2_{XZ})^{1/2}}{192 \pi r_{XZ}^4},
\end{equation}
where $r_{XZ} \equiv m_X/m_Z$.

After integrating our the $Z$, this model yields the following effective interaction for elastic scattering with nuclei (retaining only unsuppressed terms):
\begin{equation}
\mathcal{L}_{\rm eff} \supset  \frac{i g_{\chi} g_{qv}}{m^2_Z}   \left(X_{\nu} \partial_\mu X^{\nu \dagger} \bar{q} \gamma^{\mu} q + \text{h.c.} \right).
\end{equation}
In the non-relativistic limit, this yields the following WIMP-nucleus cross section:
\begin{eqnarray}
\sigma_{\chi N} &=& \frac{g^2_{\chi} \mu^2_{\chi N}}{\pi m^4_{Z}} \bigg[Z(2 g_{uv}+g_{dv}) +(A-Z)(g_{uv}+2g_{dv})\bigg]^2 \nonumber \\
&\approx& \frac{g^2_{\chi} (g_1^2+g_2^2)\mu^2_{\chi N}}{16 \pi m^4_{Z}} (A-Z)^2,
\end{eqnarray}
where $Z$ and $A$ are the atomic number and atomic mass of the target nucleus, and $\mu_{\chi N}$ is the reduced mass of the dark matter-nucleus system.

In the right frame of Fig.~\ref{fig:CS}, we see that this combination of constraints from direct detection experiments and the invisible width of the $Z$ strongly rules out the entire parameter space of this model.

%%%%%%%%%%%%%%%%%%%%%%%%%%%%%%%%%%%%%%%%%%%%%%%%%%%%%%
\section{Higgs Mediated Dark Matter}
\label{H}

%%%%%%%%%%%%%%%%%%%%%%%%%%%%%%%%%%%%%%%%%%%%%%%%%%%%%%
\subsection{Fermionic dark matter}
%%%%%%%%%%%%%%%%%%%%%%%%%%%%%%%%%%%%%%%%%%%%%%%%%%%%%%

In this subsection, we consider a dark matter candidate that is either a Dirac or Majorana fermion, with the following interactions with the SM Higgs boson: 
\begin{equation}
\mathcal{L} \supset  \left[a \bar{\chi}( \lambda_{\chi s} + \lambda_{\chi p} i \gamma^5 ) \chi \right] H,  
\end{equation}
where once again $a=1 (1/2)$ in the Dirac (Majorana) case. The quantities $\lambda_{\chi s}$ and $\lambda_{\chi p}$ denote the scalar and pseudoscalar couplings between the dark matter and the SM Higgs, respectively.

\begin{figure*}
\begin{center}
\begin{tabular}{cc}
    \includegraphics[width=0.48\textwidth]{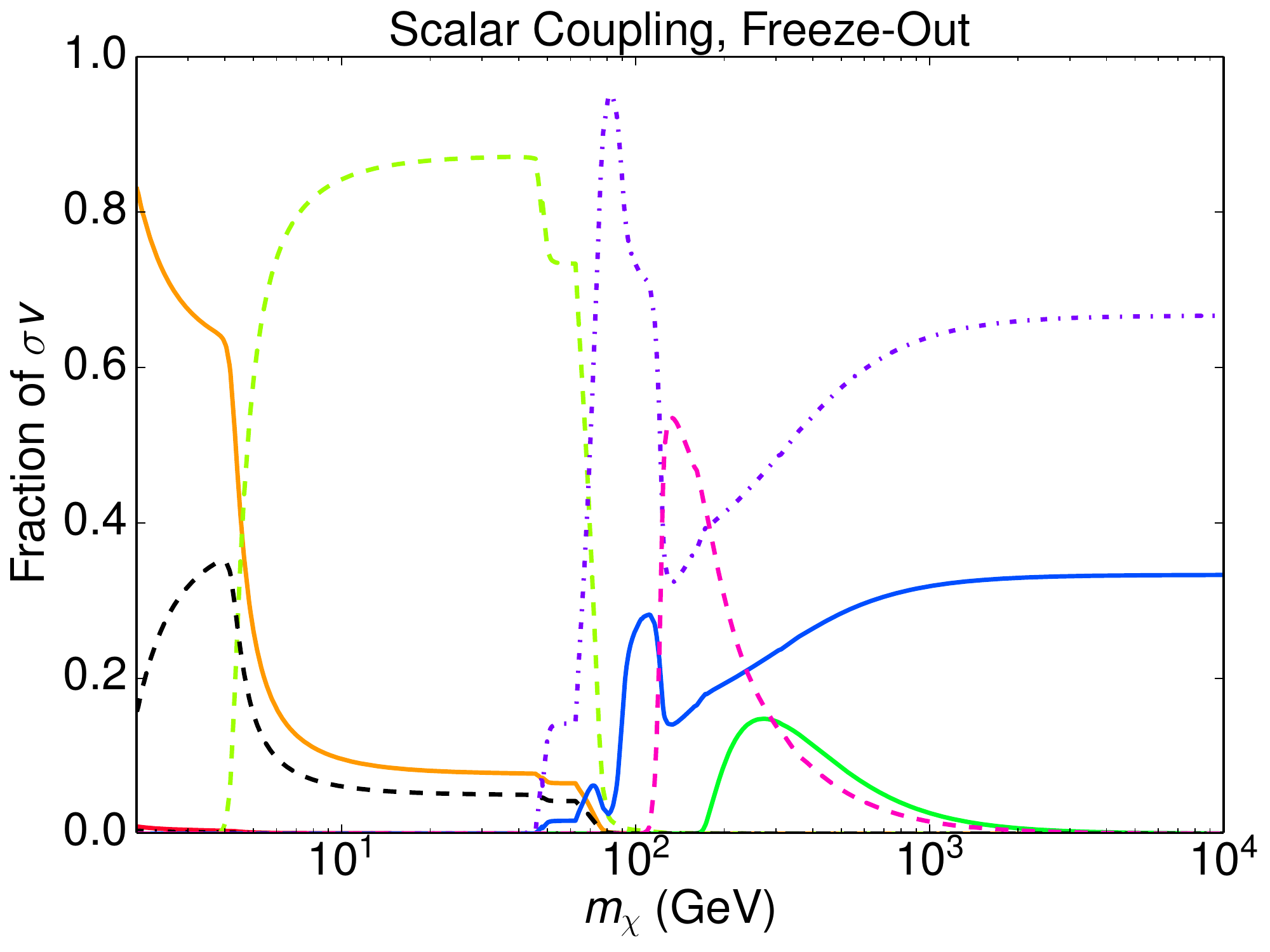} &   \includegraphics[width=0.48\textwidth]{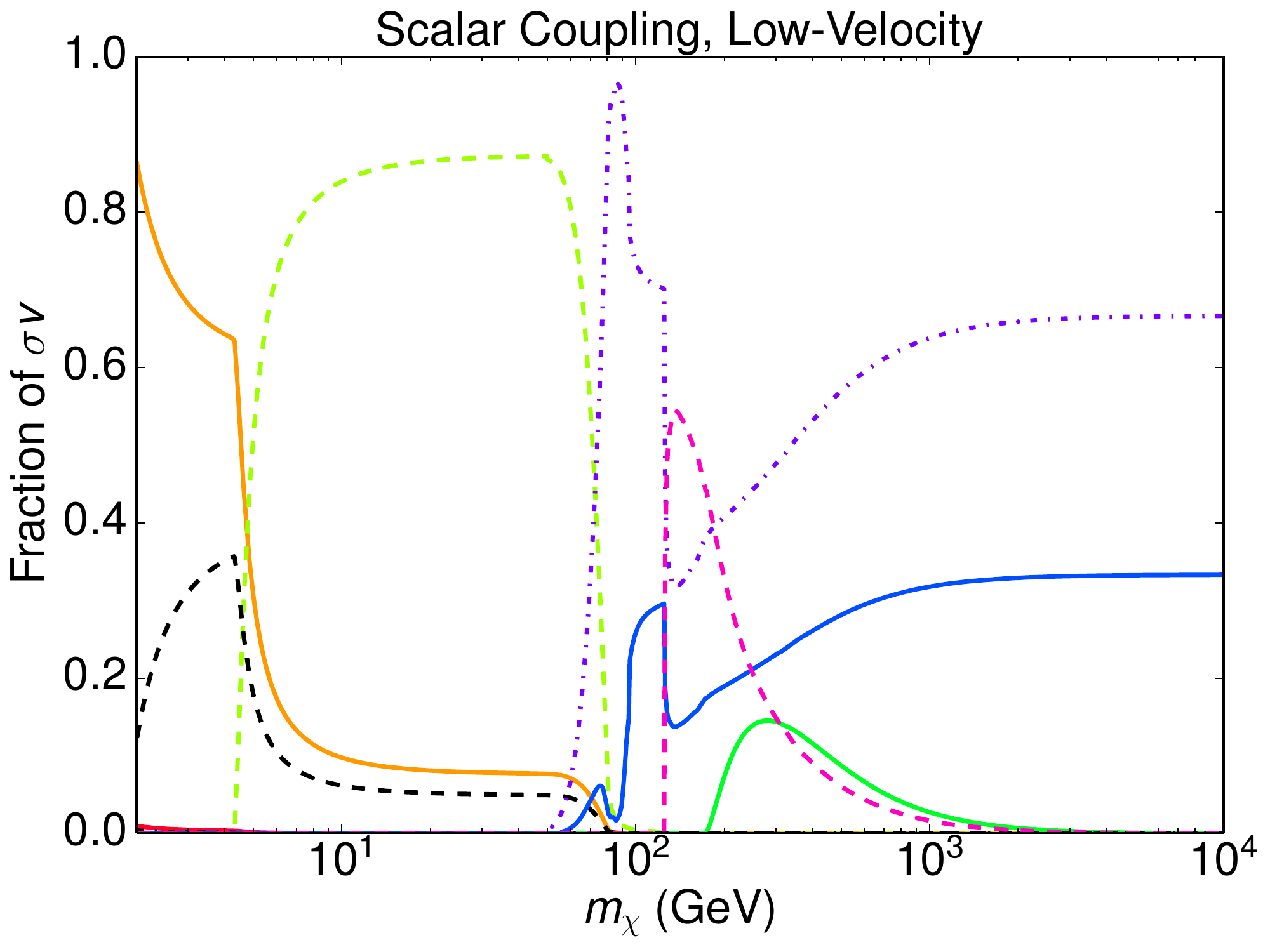} \\
        \includegraphics[width=0.48\textwidth]{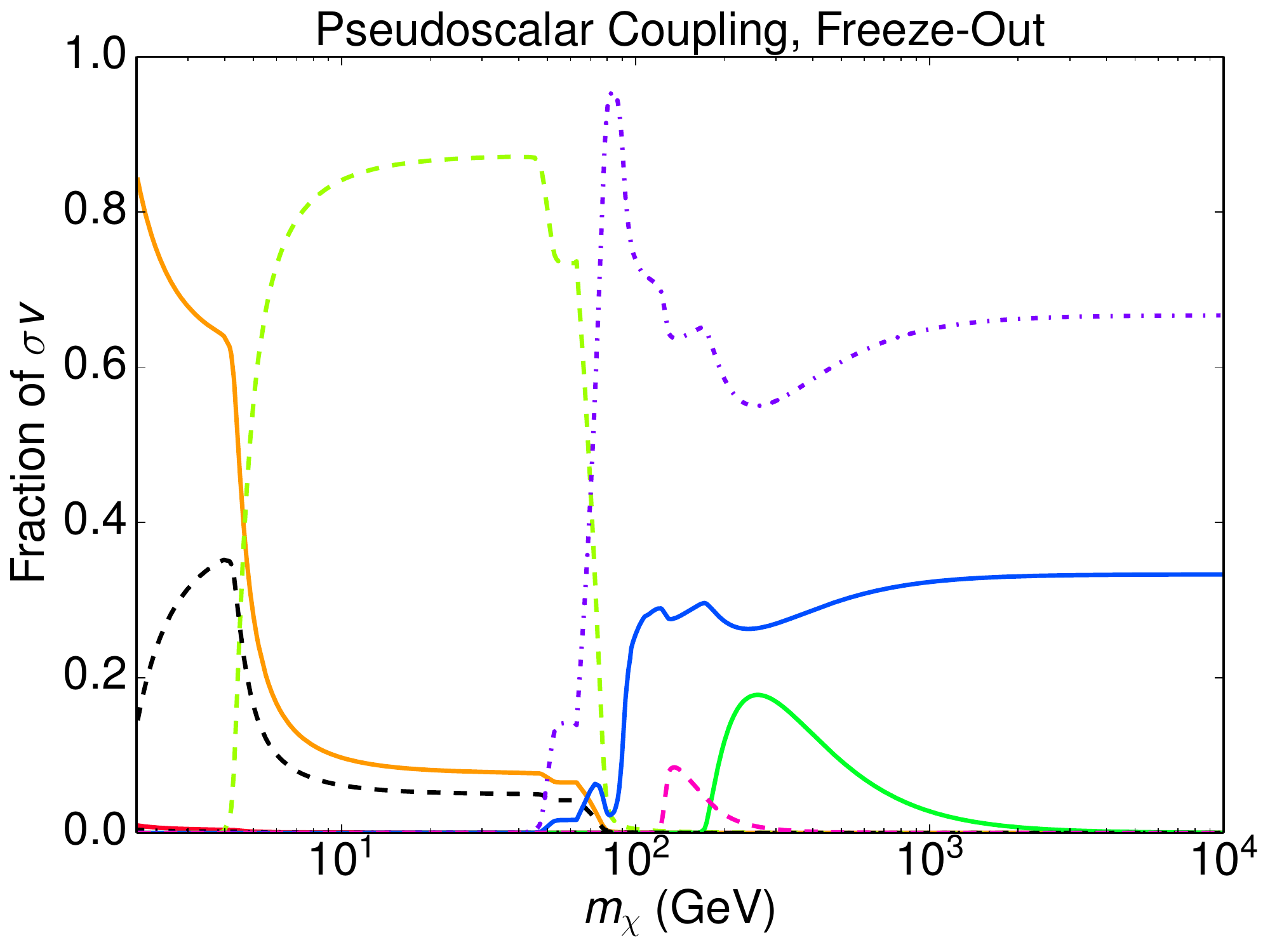} &   \includegraphics[width=0.48\textwidth]{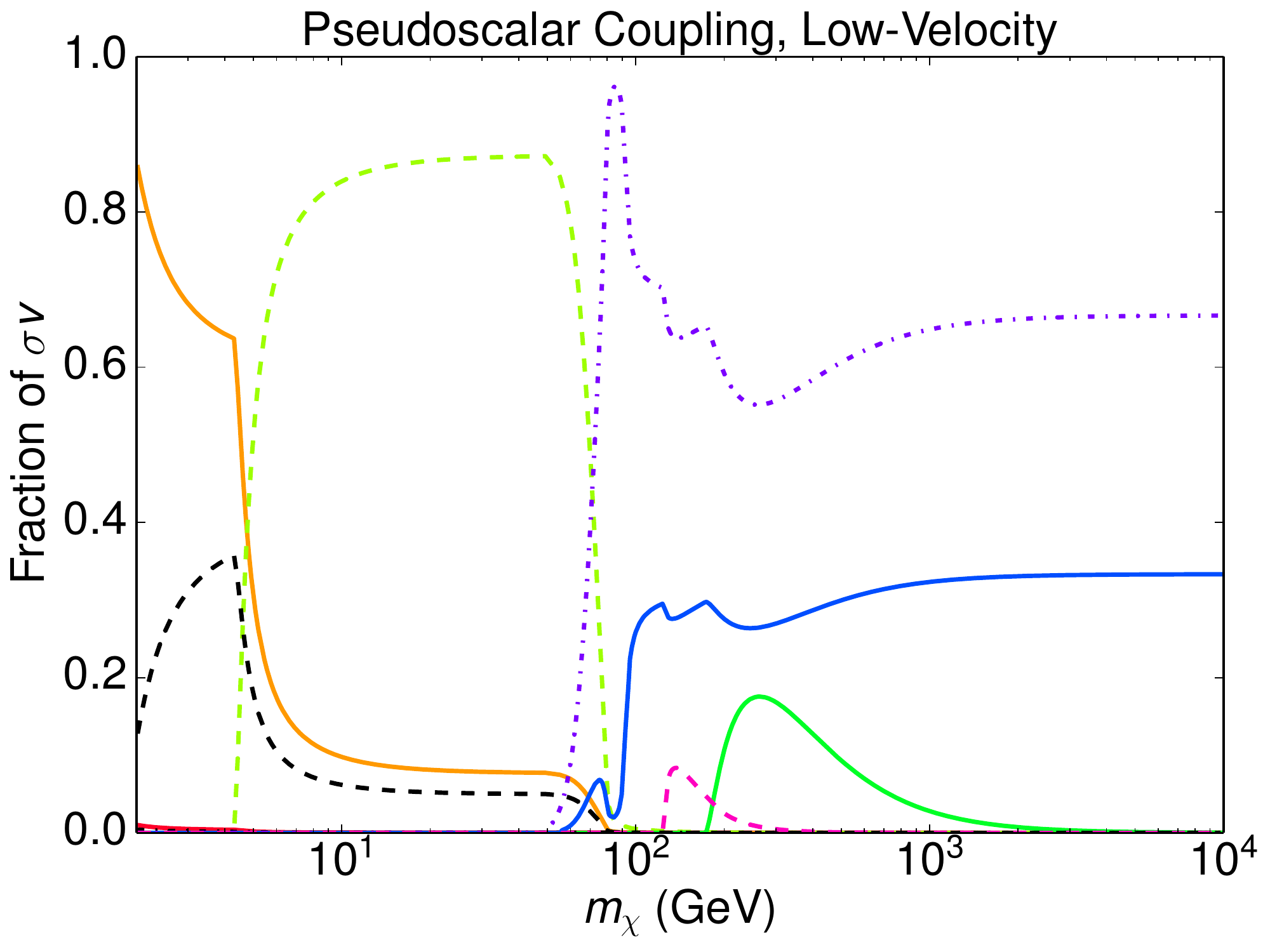} 
    \end{tabular}
    \includegraphics[width=0.98\textwidth]{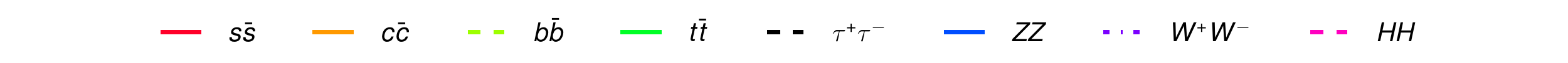} 
\caption[]{The fraction of dark matter annihilations that proceed to each final state, as evaluated at the temperature of thermal freeze-out (left) and at $v=10^{-3}\,c$, as is typically relevant for indirect detection (right). The upper and lower frames correspond to dark matter in the form of a fermion with purely scalar or purely pseudoscalar couplings to the Standard Model Higgs boson, respectively.}
\label{fig:scalar}
\end{center}
\end{figure*}

Dark matter annihilations in this model depend strongly on the choice of scalar or pseudoscalar couplings. In particular, scalar couplings lead to an annihilation cross section that is suppressed by two powers of velocity, whereas pseudoscalar couplings generate an $s$-wave amplitude with no such suppression. In both cases, annihilations proceed dominantly to heavy final states (see Fig.~\ref{fig:scalar}), due to the couplings of the Higgs to the particle content of the SM.

The contribution to the invisible Higgs width in this case is given by: 
\begin{equation}
\Gamma (H\to \chi\bar{\chi})  = \frac{a\, m_H }{8  \pi} \left[\lambda_{\chi p}^2 + \lambda_{\chi s}^2 \left(1- \frac{4 m_\chi^2}{m_H^2} \right) \right] \sqrt{1- \frac{ 4 m_\chi^2}{m_H^2}}.
\end{equation}
The current experimental constraint on the invisible branching fraction of the Higgs is $\Gamma_{\rm inv}/(\Gamma_{\rm inv}+\Gamma_{SM})< 0.24$,\footnote{This is derived from a combination of Run I and 2015 LHC data. See, for example, page 25 of the talk ``Search for invisible decays of the 125 GeV Higgs boson using the CMS detector'', by Nicholas Wardle,~\url{http://indico.cern.ch/event/432527/contributions/1071583/attachments/1320936/1980904/nckw_ICHEP_2016_hinv_cms.pdf}.} which for  $\Gamma^H_{SM} \approx 4.07 \, {\rm MeV}$ corresponds to the following:
\begin{equation}
\Gamma(H\to \chi \bar{\chi}) < \Gamma^H_{SM}\frac{{\rm BR}(H\to{\rm inv})}{1-{\rm BR}(H\to{\rm inv})} \approx 1.29 \, {\rm MeV}.
\end{equation}

Elastic scattering between dark matter and nuclei is entirely spin-independent in this case, with a cross section given as follows:
\begin{eqnarray}
\sigma_{\chi N} \approx \frac{\mu_{\chi N}^2}{\pi  m_H^4}  \left[Z f_p+(A-Z) f_n\right]^2 \bigg[\lambda^2_{\chi s}  + \lambda^2_{\chi p}\frac{q^2}{4m^2_{\chi}}\bigg],  
\end{eqnarray}
where $q$ is the momentum exchanged in the collision.

\begin{figure*}[t]
\begin{center}\begin{tabular}{cc}
     \includegraphics[width=0.48\textwidth]{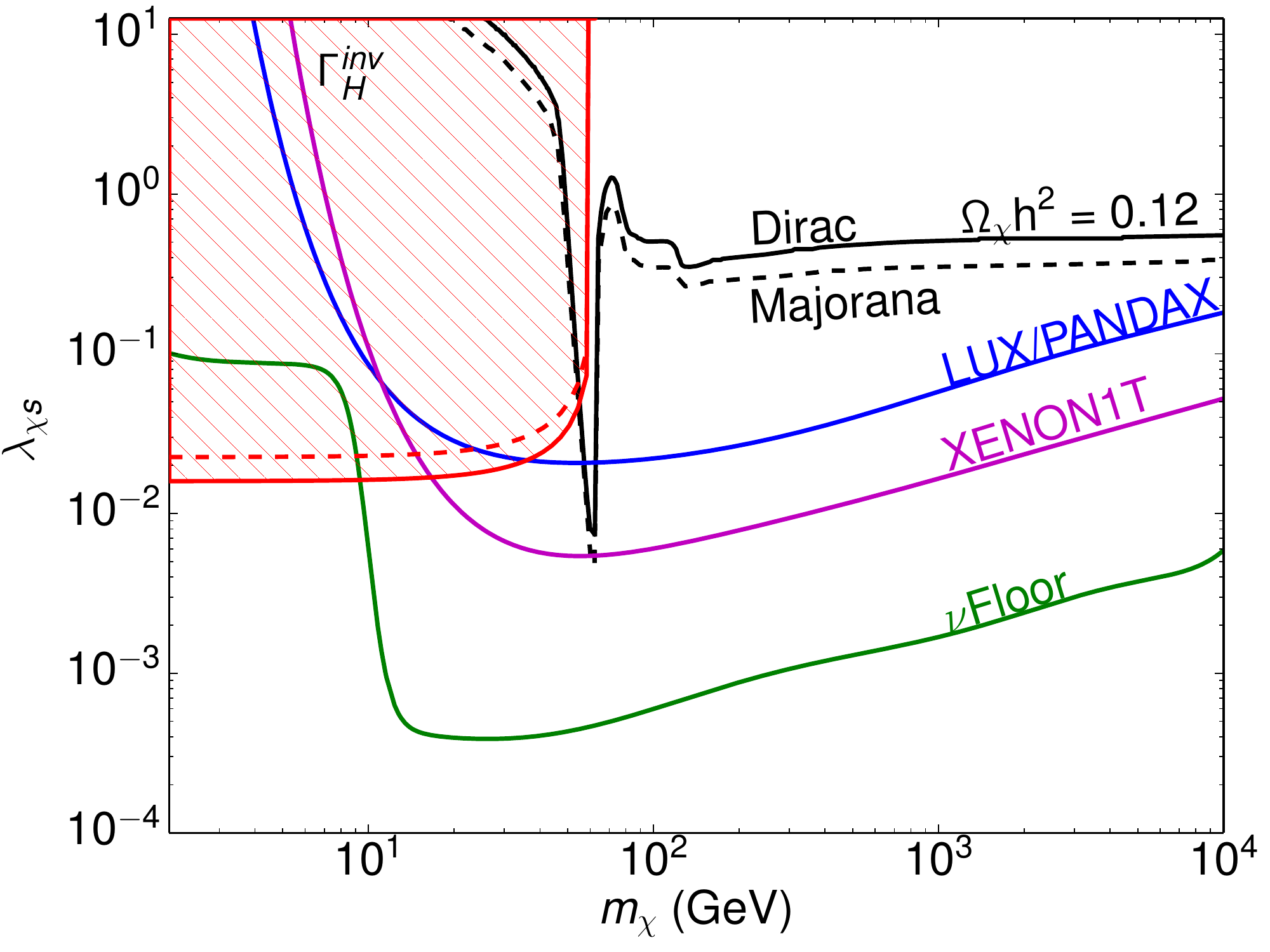} 
       \includegraphics[width=0.48\textwidth]{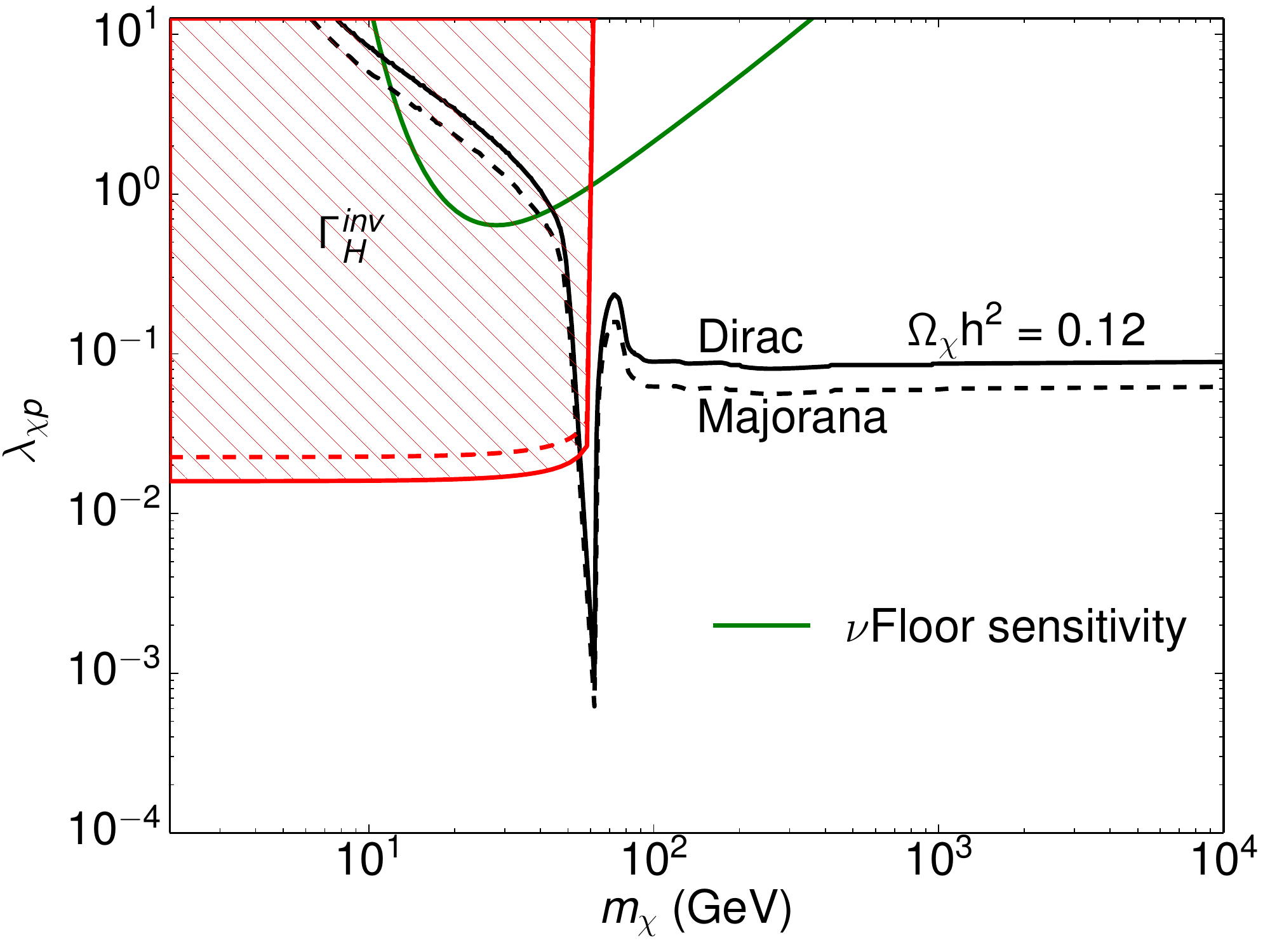} 
\end{tabular}
\caption[]{Constraints on mass and couplings of a fermionic dark matter candidate which annihilates through the Standard Model Higgs boson. The solid black contours indicate the value of the coupling for which the thermal relic abundance matches the measured cosmological dark matter density, $\Omega_{\chi} h^2 = 0.12$. The shaded regions are excluded by measurements of the invisible Higgs width. The left and right frames depict the cases of a purely scalar or pseudoscalar coupling between the dark matter and the Higgs, respectively. In the scalar case, the vast majority of this parameter space is excluded by the current constraints from LUX and PandaX-II~\cite{Akerib:2016vxi,Tan:2016zwf}. The only currently viable region ($m_{\chi} =$ 56-62 GeV is expected to be probed in the near future by XENON1T~\cite{Aprile:2015uzo}. Due to the momentum suppression of the elastic scattering cross section, the case of dark matter with a pseudoscalar coupling to the Higgs is much less strongly constrained.}
\label{fig:Scalar}
\end{center}
\end{figure*}

In Fig.~\ref{fig:Scalar}, we summarize the constraints on this scenario. In the case of a purely scalar coupling ($\lambda_{\chi p}=0$, shown in the left frame), the combination of the invisible Higgs width measurement and the results of direct detection experiments rule out nearly all of the parameter space. The exception is the mass range within a few GeV of the Higgs pole, $m_{\chi} =$ 56-62 GeV. In this case, future experiments such as XENON1T are expected to test the remaining region of parameter space.

In the case of a purely pseudoscalar coupling ($\lambda_{\chi s}=0$, shown in the right frame of Fig.~\ref{fig:Scalar}), the momentum suppression of the elastic scattering cross section strongly reduces the prospects for direct detection experiments, earning this scenario the moniker of ``coy dark matter''~\cite{Boehm:2014hva,Hektor:2014kga,Kozaczuk:2015bea}. Naively, we expect the sensitivity of direct detection experiments to the coupling, $\lambda_{\chi p}$, to be suppressed relative to $\lambda_{\chi s}$ by a factor of $q/2m_{\chi}$, which for typical scattering events is on the order of $10^{-3}$. Simply rescaling the results shown in the left frame of Fig.~\ref{fig:Scalar} by this factor leads us to conclude that current (LUX, PandaX-II) and near future (XENON1T) experiments will not be sensitive to dark matter in this scenario. It is less clear, however, whether a larger experiment, with a sensitivity to cross sections near the neutrino floor, might be sensitive to this scenario. With this in mind, we have calculated the sensitivity of such an experiment to a dark matter candidate with a momentum suppressed elastic scattering cross section with nuclei.

Dark matter with velocity or momentum suppressed scattering has been considered previously in the literature (see, for example, Refs.~\cite{Chang:2009yt,Feldstein:2009tr,Chang:2008xa,Dienes:2013xya,Fan:2010gt,Dent:2016iht,Beniwal:2015sdl,Gluscevic:2015sqa,Kumar:2013iva,Savage:2008er}). To compute the number of events in a large volume xenon experiment, we follow the procedure outlined in Ref.~\cite{Savage:2008er}, adopting a standard Maxwellian velocity distribution ($v_0 = 220$ km/s, $v_{\rm esc} = 544$ km/s, $\bar{v}_{\rm Earth}=245$ km/s), a local density of $0.3 \, {\rm GeV/cm^3}$, and a Helm form factor~\cite{Lewin:1995rx}. Regarding the detector specifications, we assume an optimistic scenario with an energy independent efficiency of 25\% and perfect energy resolution. We consider events with nuclear recoil energies between 6 and 30 keV, where this lower limits was imposed in order to reduce the rate of neutrino-induced background events~\cite{Billard:2013qya,Akerib:2015cja}. From the calculated event rate, we apply Poisson statistics to place a 90\% confidence level constraint on the dark matter coupling, assuming that zero events are observed. In the right frame of Fig.~\ref{fig:Scalar}, we plot the projected constraint from such an experiment after collecting an exposure of 30 ton-years, which is approximately the exposure that we estimate will accumulate between $\sim$1-3 neutrino-induced background events. From this, we conclude that even with such an idealized detector, it will not be possible to test a dark matter candidate with a purely pseudoscalar coupling to the Higgs.

\begin{figure*}[t]
\begin{center}\begin{tabular}{cc}
    \includegraphics[width=0.48\textwidth]{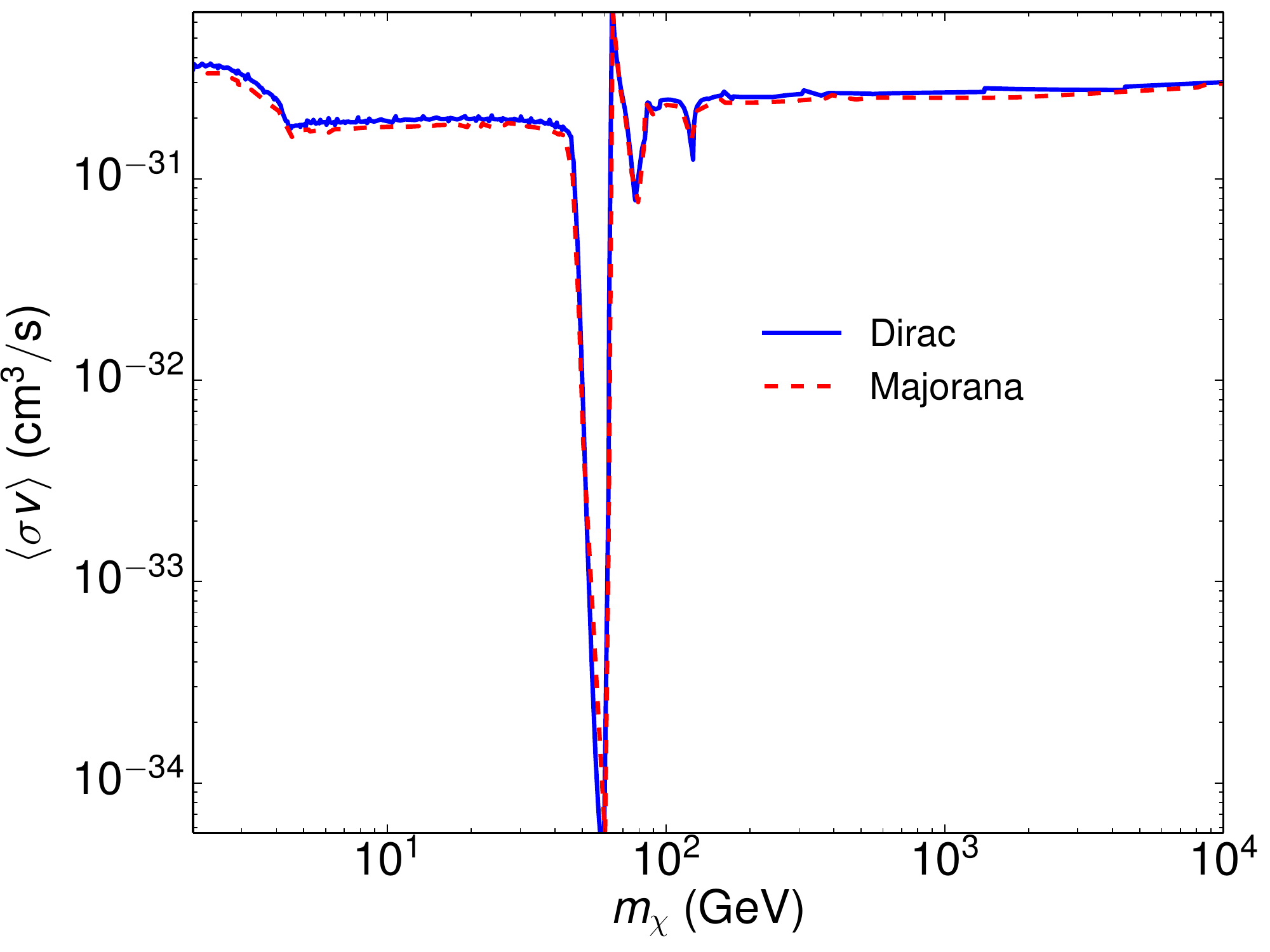}
    \includegraphics[width=0.48\textwidth]{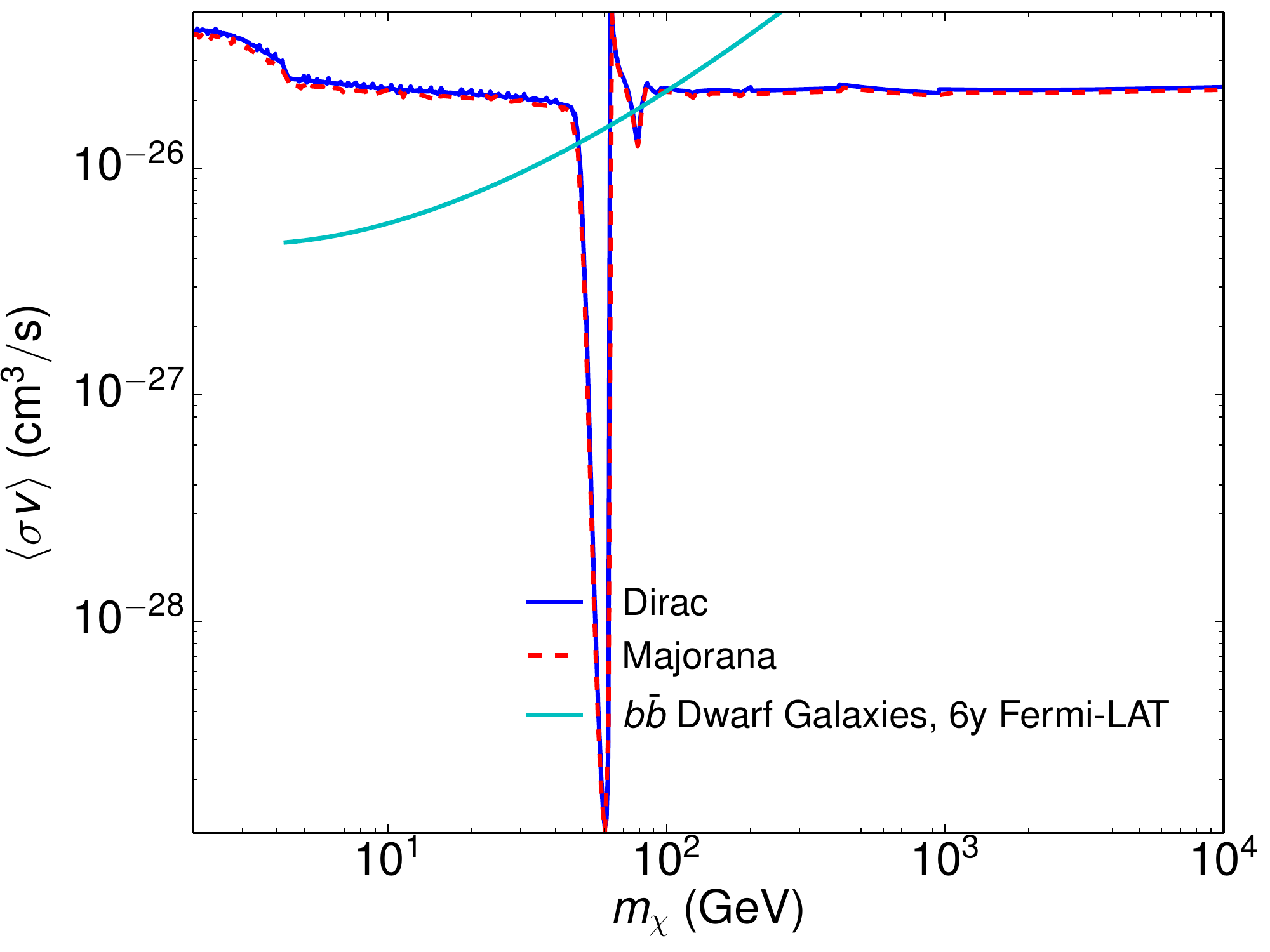}
\end{tabular}
\caption[]{The effective low-velocity annihilation cross section (relevant for indirect detection) for Dirac or Majorana dark matter with a scalar (left) or pseudoscalar (right) coupling to the SM Higgs boson. For the case of scalar couplings, the cross section is always well below the sensitivity of Fermi and other existing indirect detection experiments. In the pseudoscalar case, the prospects for indirect detection are much more encouraging. In the right frame, we also show the current constraint from Fermi's observation of dwarf spheroidal galaxies~\cite{Ackermann:2015zua}.}
\label{fig:PScalar}
\end{center}
\end{figure*}

In the case of dark matter with a scalar coupling and near the Higgs pole, the low-velocity annihilation cross section is suppressed by two powers of velocity, making such a scenario well beyond the reach of any planned or proposed indirect detection experiment (see the left frame of Fig.~\ref{fig:PScalar}). In the case of dark matter with a pseudoscalar coupling to the Higgs, however, the low-velocity annihilation rate is unsuppressed, leading to more promising prospects for indirect detection (for analytic expressions of these cross section, see the Appendix of Ref.~\cite{Berlin:2014tja}). In the right frame of Fig.~\ref{fig:PScalar}, we plot the low-velocity annihilation cross section (as relevant for indirect detection) for fermionic (Dirac or Majorana) dark matter with a pseudoscalar coupling to the SM Higgs boson. In this case, constraints from Fermi's observations of dwarf spheroidal galaxies~\cite{Ackermann:2015zua} may be relevant, depending on the precise value of the dark matter mass. We also note that uncertainties associated with the distribution of dark matter in these systems could plausibly weaken these constraints to some degree~\cite{Bonnivard:2014kza,Bonnivard:2015vua,Ichikawa:2016nbi,Klop:2016lug}. It may also be possible in this scenario~\cite{Berlin:2015wwa,Boehm:2014hva,Ipek:2014gua,Fan:2015sza,Kim:2016csm} to generate the gamma-ray excess observed from the region surrounding the Galactic Center~\cite{Daylan:2014rsa,Hooper:2010mq,Hooper:2011ti,Goodenough:2009gk,Abazajian:2012pn,Gordon:2013vta,TheFermi-LAT:2015kwa}.

%%%%%%%%%%%%%%%%%%%%%%%%%%%%%%%%%%%%%%%%%%%%%%%%%%%%%%
\subsection{Scalar dark matter}
%%%%%%%%%%%%%%%%%%%%%%%%%%%%%%%%%%%%%%%%%%%%%%%%%%%%%%

In the case of scalar dark matter with a coupling to the SM Higgs boson, we consider a Higgs Portal interaction, described by the following Lagrangian:
\begin{equation}
\mathcal{L} \supset  a\, \lambda_{\phi H}\,\bigg[ v  H \phi ^2   + \frac{1}{2} H^2 \phi^2 \bigg],
\end{equation}
where $a=1 \,(1/2)$ in the case of a complex (real) scalar, and $v$ is the vacuum expectation value of the SM Higgs boson. 
 
\begin{figure*}[t]
\begin{center}\begin{tabular}{cc}
      \includegraphics[width=0.48\textwidth]{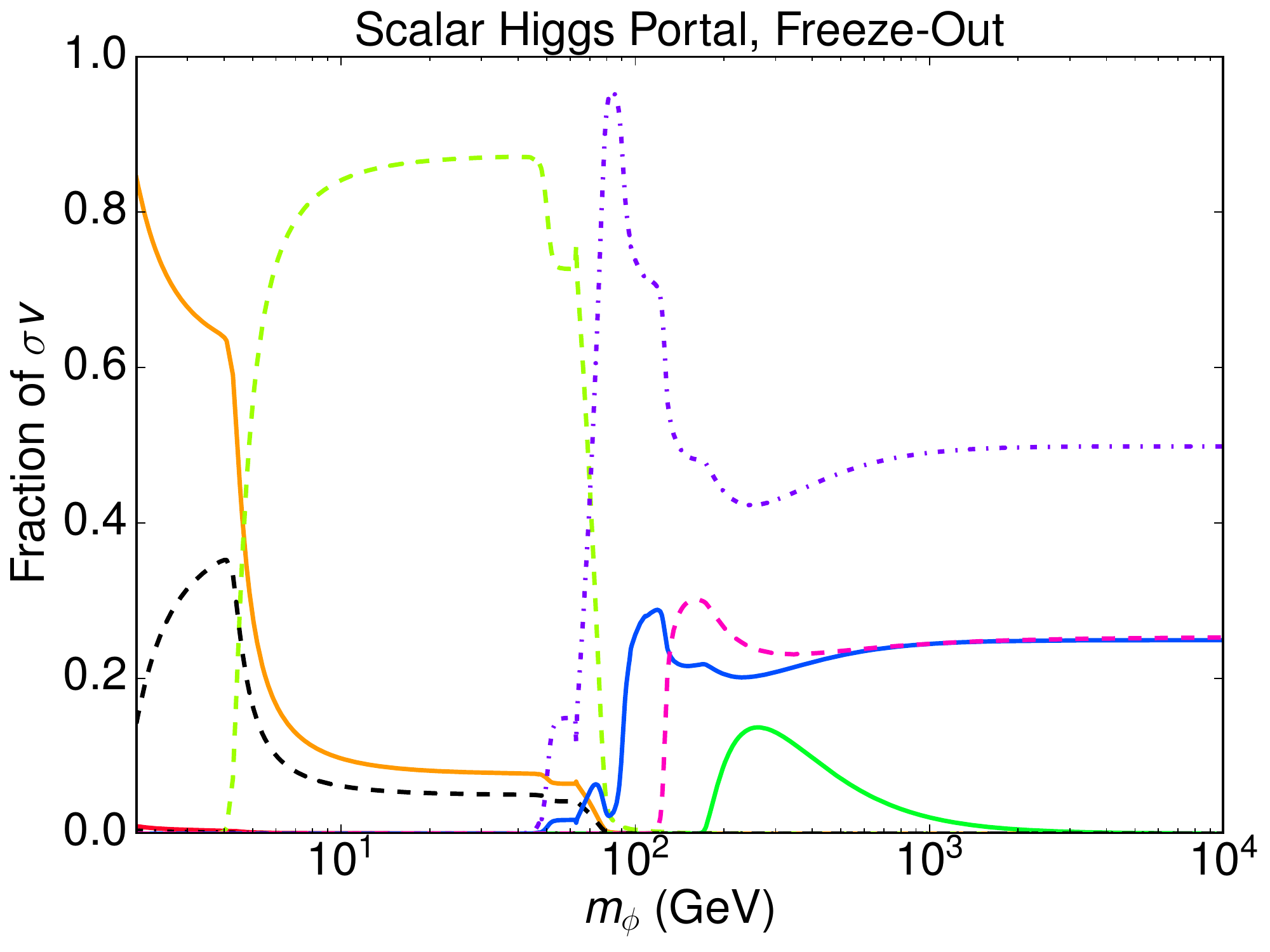}      
      \includegraphics[width=0.48\textwidth]{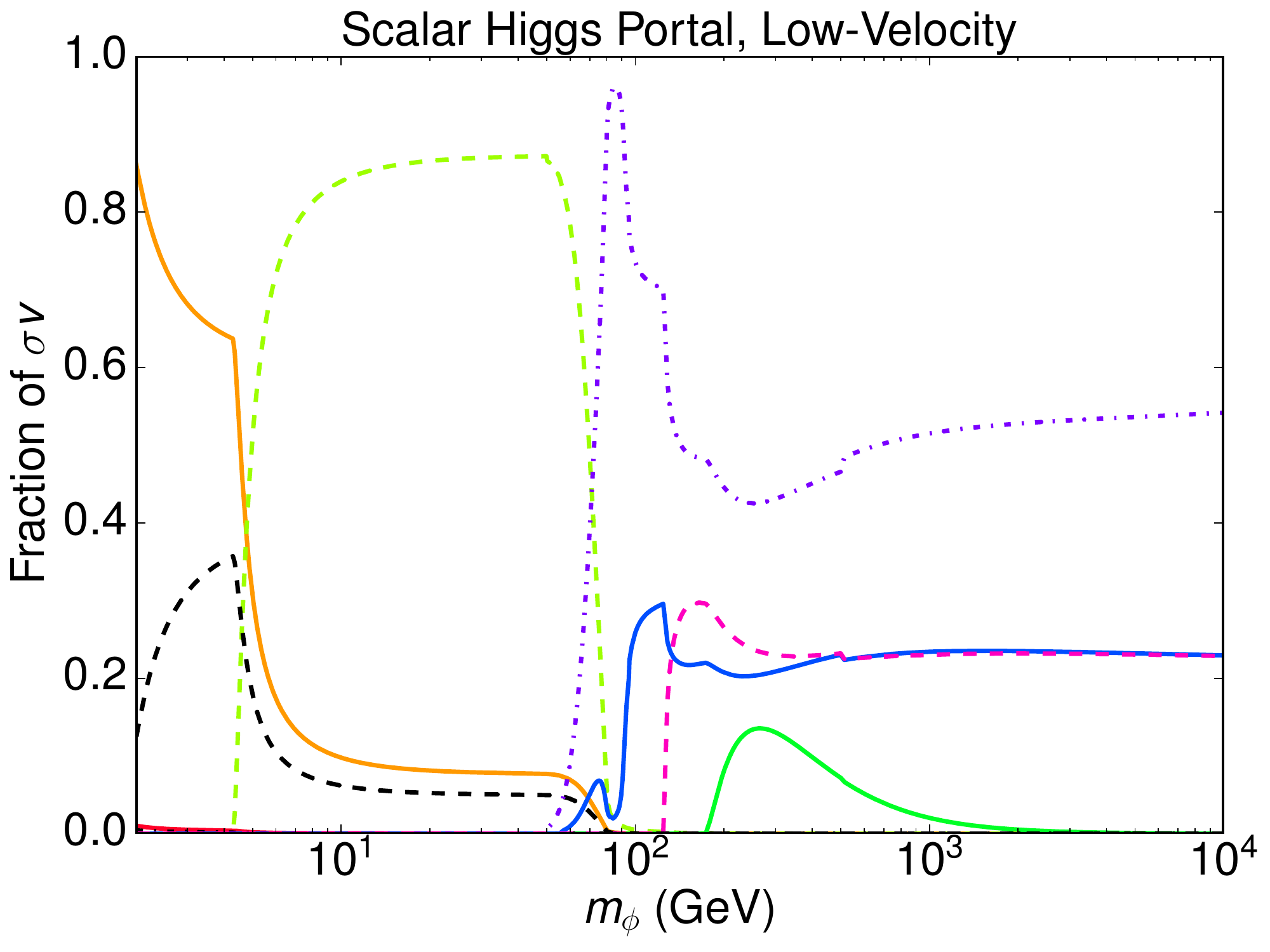} 
 \end{tabular}
     \includegraphics[width=0.98\textwidth]{figures/H_legend.pdf} 
\caption[]{The fraction of dark matter annihilations that proceed to each final state, as evaluated at the temperature of thermal freeze-out (left) and at $v=10^{-3}\,c$, as is typically relevant for indirect detection (right), for the case of scalar dark matter coupled to the Standard Model Higgs boson.}
\label{fig:S1}
\end{center}
\end{figure*}

In this class of models, the dark matter annihilates without velocity suppression, and preferentially to heavy final states (see Fig.~\ref{fig:S1}). The contribution to the invisible Higgs width in this case is given by: 
\begin{eqnarray}
\Gamma (H\to \phi \phi^\dagger)  &=&  \frac{a\, v^2 \lambda_{\phi H}^2 }{16  \, \pi  m_H}  \, \sqrt{1-\frac{4m_\phi^2}{m_H^2}}.
\end{eqnarray}

\begin{figure*}[t]
\begin{center}\begin{tabular}{cc}
    \includegraphics[width=0.49\textwidth]{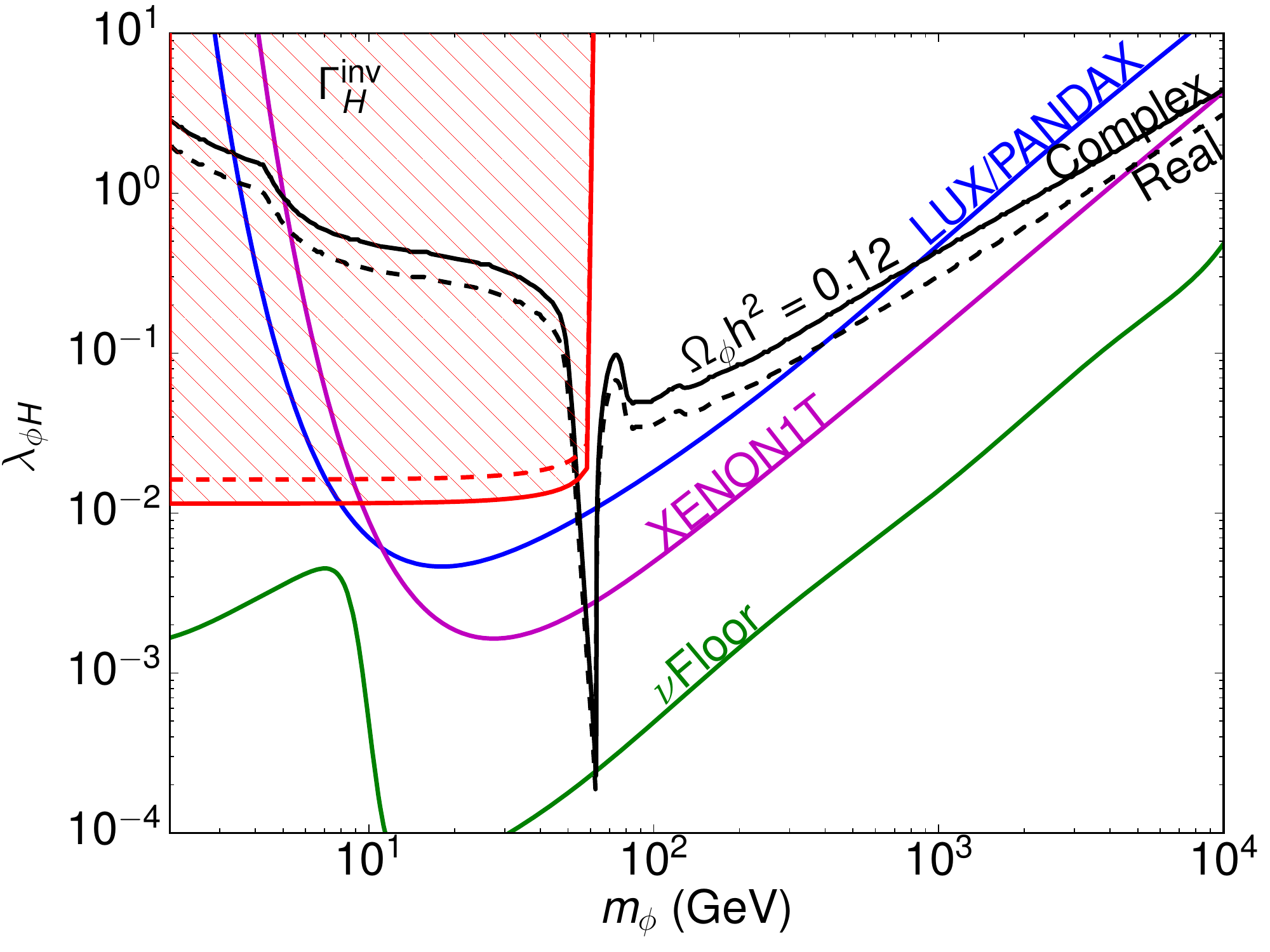}
     \end{tabular}
\caption[]{Constraints on mass and couplings of a complex scalar dark matter candidate which annihilates through a Higgs portal coupling. The solid black contour indicates the value of the coupling for which the thermal relic abundance matches the measured cosmological dark matter density, $\Omega_{\chi} h^2 = 0.12$. The shaded region is excluded by measurements of the invisible Higgs width, and the region above the solid blue line is excluded by the current constraints from LUX and PandaX-II~\cite{Akerib:2016vxi,Tan:2016zwf}. This scenario is currently viable only if the mass of the dark matter candidate is near the Higgs pole ($m\simeq m_H/2$) or if $m_{\phi} \gsim 400$ GeV.}
\label{fig:contributt}
\end{center}
\end{figure*}

In Fig.~\ref{fig:contributt}, we plot a summary of the constraints in this class of models. In this case, we find that complex (real) scalar dark matter with a mass greater than 840 GeV (400 GeV) is not currently constrained, along with the region near the Higgs pole. XENON1T is expected to probe the remaining high mass window up to 10 TeV (5 TeV). Similar constraints can be found in the recent analysis of~\cite{He:2016mls}.

\begin{figure*}[t]
\begin{center}\begin{tabular}{cc}
        \includegraphics[width=0.48\textwidth]{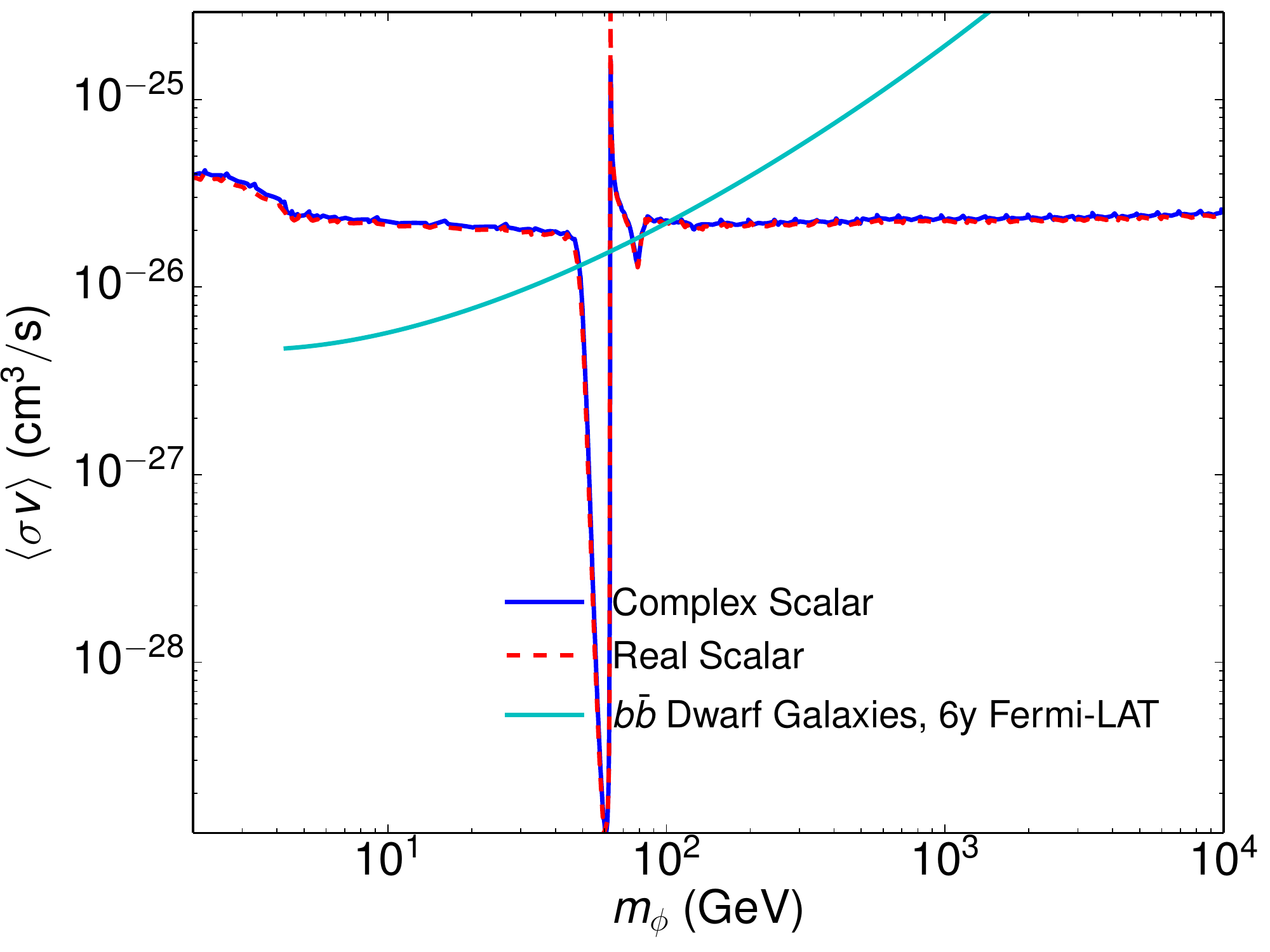}
\end{tabular}
\caption[]{The effective low-velocity annihilation cross section (relevant for indirect detection) for complex or real scalar dark matter with a Higgs portal coupling. We also show the current constraint from Fermi's observation of dwarf spheroidal galaxies~\cite{Ackermann:2015zua}.}
\label{fig:S3}
\end{center}
\end{figure*}

In Fig.~\ref{fig:S3}, we plot the low-velocity annihilation cross section (as relevant for indirect detection) for scalar dark matter with a Higgs portal coupling. In the currently allowed mass range near the Higgs pole, this class of models predicts a very small low-velocity annihilation cross section, which is likely unable to generate the measured intensity of the Galactic Center gamma-ray excess~\cite{Daylan:2014rsa,Hooper:2010mq,Hooper:2011ti,Goodenough:2009gk,Abazajian:2012pn,Gordon:2013vta,TheFermi-LAT:2015kwa} (see also, Refs.~\cite{Cuoco:2016jqt,Sage:2016xkb}).

%%%%%%%%%%%%%%%%%%%%%%%%%%%%%%%%%%%%%%%%%%%%%%%%%%%%%%
\subsection{Vector dark matter}
%%%%%%%%%%%%%%%%%%%%%%%%%%%%%%%%%%%%%%%%%%%%%%%%%%%%%%

\begin{figure*}[t]
\begin{center}\begin{tabular}{cc}
    \includegraphics[width=0.48\textwidth]{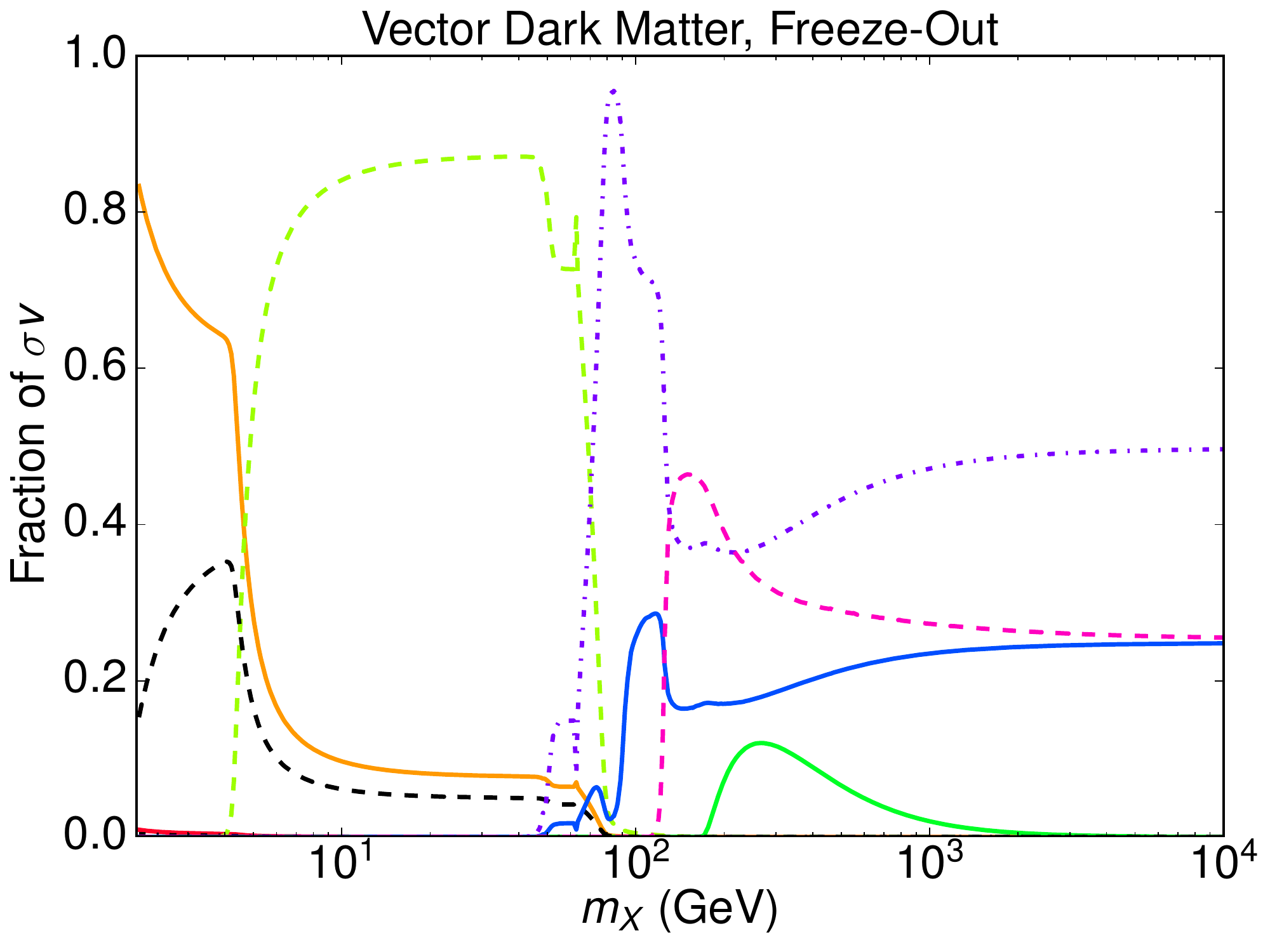} &   \includegraphics[width=0.48\textwidth]{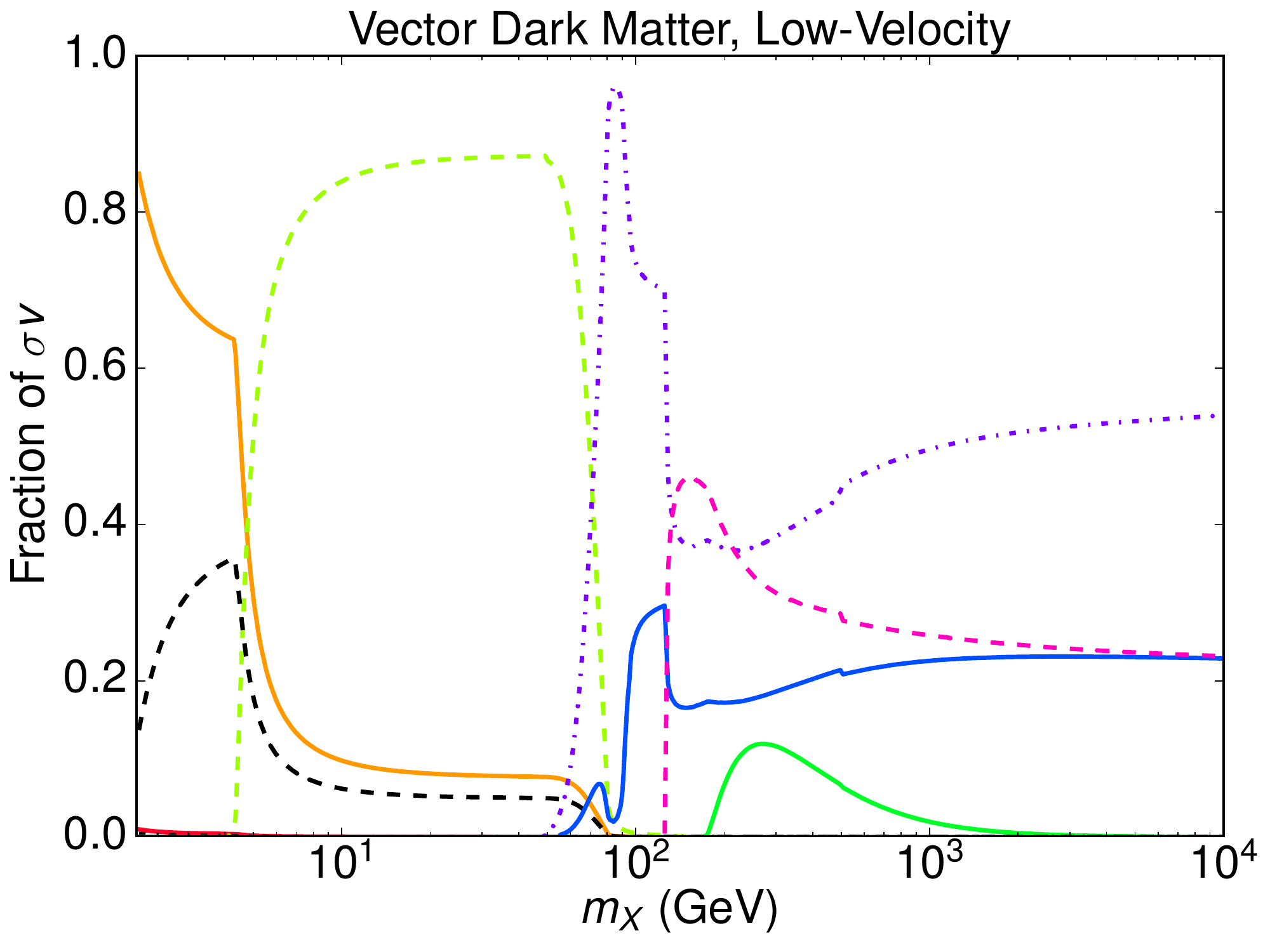} 
        \end{tabular}
            \includegraphics[width=0.98\textwidth]{figures/H_legend.pdf} 
\caption[]{The fraction of dark matter annihilations that proceed to each final state, as evaluated at the temperature of thermal freeze-out (left) and at $v=10^{-3}\,c$, as is typically relevant for indirect detection (right), for the case of dark matter in the form of a vector which annihilates through its coupling to the Standard Model Higgs boson.}
\label{fig:V1}
\end{center}
\end{figure*}

In the case of vector dark matter, we again consider a Higgs Portal interaction, which is described in this case by the following Lagrangian:

\begin{equation}
\mathcal{L} \supset a \lambda_{X H}  \left[ v H X^\mu X_\mu^\dagger
+\frac{1}{2} H^2 X^\mu X_\mu^\dagger \right],
\end{equation}
where $a=1 \,(1/2)$ in the case of a complex (real) vector. 
As in the cases considered in the previous subsection, dark matter annihilates without velocity suppression in this class of models, and preferentially to heavy final states (see Fig.~\ref{fig:V1}).

The contribution to the invisible Higgs width is given in this model by:
\begin{equation}
\Gamma (H\to X X^\dagger)  = \frac{a \, \lambda_{X H}^2 v^2 m_H^3}{64
\pi  m_X^4} \left(1 -4 \frac{m_X^2}{m_H^2}+12
\frac{m_X^4}{m_H^4}\right)  \sqrt{1-\frac{4m_X^2}{m_H^2}}.
\end{equation}

\begin{figure*}[t]
\begin{center}\begin{tabular}{cc}
            \includegraphics[width=0.49\textwidth]{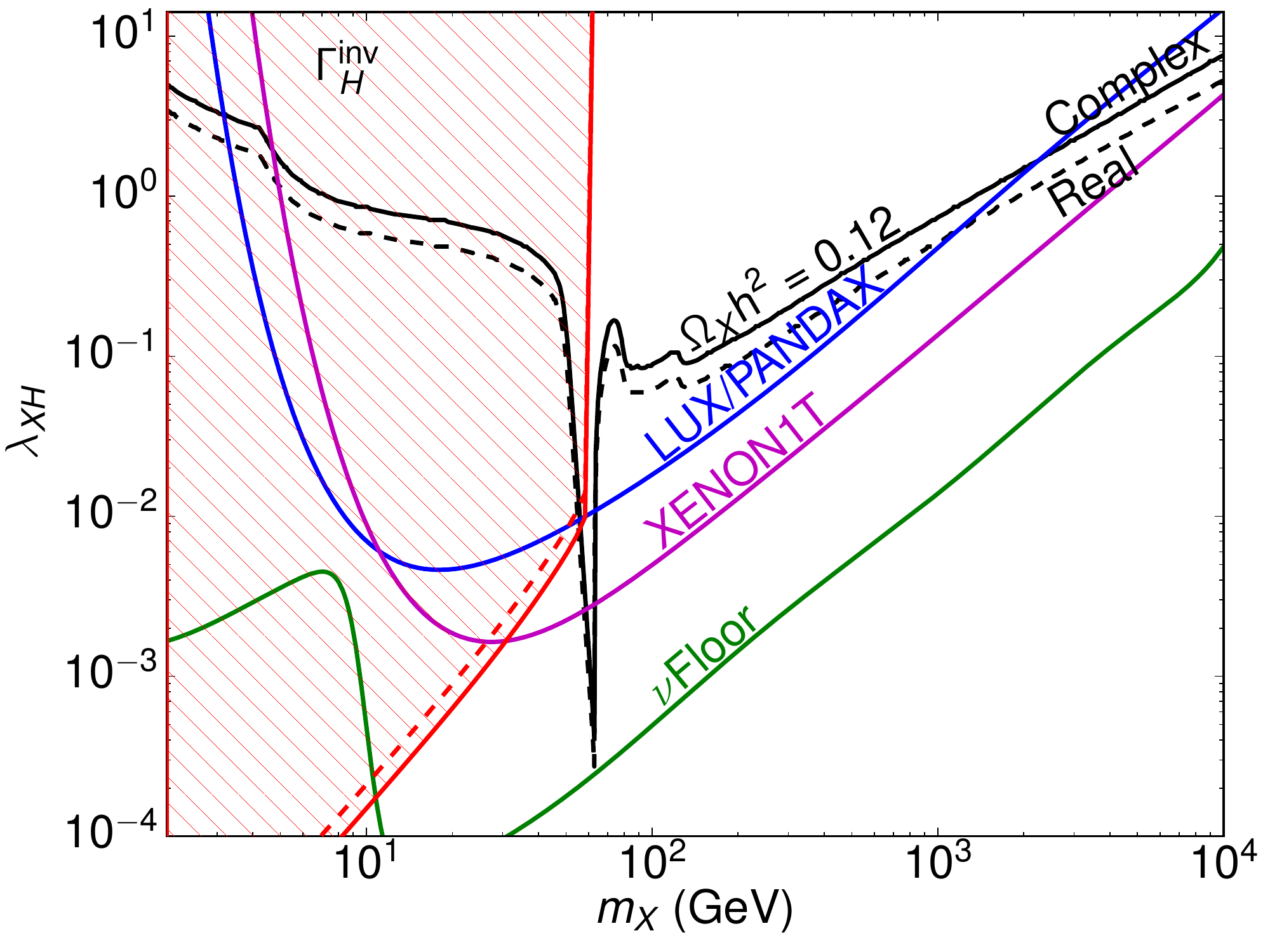} 
     \end{tabular}
\caption[]{Constraints on mass and couplings of a complex vector dark matter candidate which annihilates through the Standard Model Higgs boson. The solid black contour indicates the value of the coupling for which the thermal relic abundance matches the measured cosmological dark matter density, $\Omega_{\chi} h^2 = 0.12$. The shaded region is excluded by measurements of the invisible Higgs width, and the region above the solid blue line is excluded by the current constraints from LUX and PandaX-II~\cite{Akerib:2016vxi,Tan:2016zwf}. This scenario is currently viable only if the mass of the dark matter candidate is near the Higgs pole ($m\simeq m_H/2$) or if $m_{X} \gsim 1160$ GeV.}
\label{fig:V2}
\end{center}
\end{figure*}

The constraints on this scenario are summarized in Fig.~\ref{fig:V2}. The combination of constraints from LUX/PandaX-II and on the invisible Higgs width rule out all of the parameter space in this class of models, with the exception of the mass range near the Higgs pole, $m_{X} \simeq m_H/2$ or for $m_{X} \gsim 1160$ GeV. XENON1T is expected in the near future to probe most of this remaining high mass window, covering nearly the entire range of perturbative values for the coupling, $\lambda_{XH} \lsim 4 \pi$.

\begin{figure*}[t]
\begin{center}\begin{tabular}{cc}
      \includegraphics[width=0.48\textwidth]{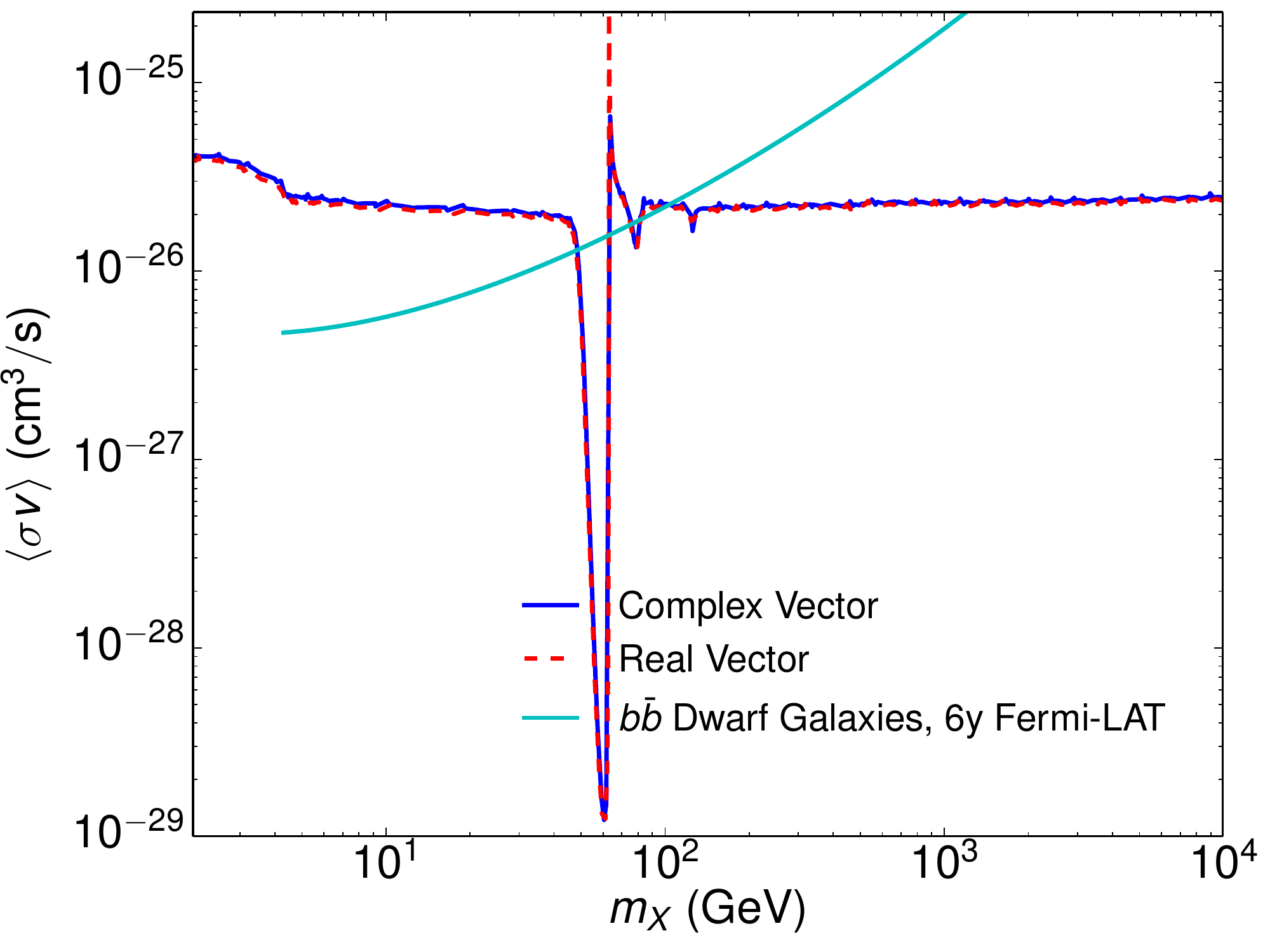}
\end{tabular}
\caption[]{The effective low-velocity annihilation cross section (relevant for indirect detection) for complex or real vector dark matter with a coupling to the Standard Model Higgs boson. We also show the current constraint from Fermi's observation of dwarf spheroidal galaxies~\cite{Ackermann:2015zua}.}
\label{fig:V3}
\end{center}
\end{figure*}

We plot in Fig.~\ref{fig:V3} the low-velocity annihilation cross section (as relevant for indirect detection) in this class of models, along with the constraints from Fermi's observations of dwarf spheroidal galaxies~\cite{Ackermann:2015zua}. 
In the currently allowed mass range near the Higgs pole, this class of models predicts a very small low-velocity annihilation cross section, which is likely unable to generate the measured intensity of the Galactic Center gamma-ray excess~\cite{Daylan:2014rsa,Hooper:2010mq,Hooper:2011ti,Goodenough:2009gk,Abazajian:2012pn,Gordon:2013vta,TheFermi-LAT:2015kwa}.

\section{Caveats}
\label{caveats}

The conclusions presented here rely on a number of assumptions that we have implicitly made throughout this study. In particular, we have assumed that the thermal history of the early universe is well described by the standard radiation-dominated picture. Departures from this simple thermal history could potentially reduce the couplings of the dark matter that are required to generate an acceptable thermal relic abundance, thereby relaxing the constraints from direct detection experiments and from measurements of the invisible $Z$ and Higgs widths. Examples of such scenarios include those in which the abundance of dark matter is depleted as a result of an out-of-equilibrium decay~\cite{Berlin:2016gtr,Berlin:2016vnh,Hooper:2013nia,Fornengo:2002db,Gelmini:2006pq,Kane:2015jia,Patwardhan:2015kga} or a period of late-time inflation~\cite{Davoudiasl:2015vba,Lyth:1995ka,Cohen:2008nb,2012PhRvD..85j3506B,2010PhRvL.105d1301B}.

We have also limited our analysis in this paper to couplings between pairs of dark matter particles and one or more $Z$ or Higgs bosons. We could instead have considered couplings between one dark matter particle, the $Z$ or Higgs, and an additional state. If the additional state is not much heavier than the dark matter itself, such a coupling could allow the dark matter to be depleted in the early universe through coannihilations~\cite{Griest:1990kh,Edsjo:1997bg}, without necessarily inducing a large elastic scattering cross section with nuclei. Generally speaking, if such coannihilations are to be efficient, such states must have a mass that is within roughly $\sim$10\% of the mass of the dark matter particle itself. 

Phenomenology of this kind can be easily realized if we consider a Dirac fermion that is split into a pair of nearly degenerate Majorana fermions by a small Majorana mass term. Following Refs.~\cite{TuckerSmith:2001hy,Cui:2009xq}, the Lagrangian in this scenario takes the following form:
\begin{eqnarray}
 \mathcal{L} &\supset&
\frac{1}{2}\bar{\Psi}_1i\gamma^{\mu}\partial_{\mu}\Psi_1 -
\frac{1}{2}(M-m_+)\bar{\Psi}_1{\Psi}_1  + \frac{1}{2}\bar{\Psi}_2i\gamma^{\mu}\partial_{\mu}\Psi_2 -
\frac{1}{2}(M+m_+)\bar{\Psi}_2{\Psi}_2 \\
&& + i\,g\,Q\,Z_{\mu}\,\bar{\Psi}_2\gamma_{\mu}\Psi_1 + \frac{1}{2}\,g\,Q\,Z_{\mu}\,\frac{\phantom{.}m_-}{M}\,\left(
\bar{\Psi}_2\gamma^{\mu}\gamma^5\Psi_2
-\bar{\Psi}_1\gamma^{\mu}\gamma^5\Psi_1\right) +
\mathcal{O}\left(\frac{m^2}{M^2}\right), \nonumber 
\end{eqnarray} 
where $\Psi_1$ and $\Psi_2$ are the two quasi-degenerate Majorana fermions, $m_{\pm} = (m_L\pm m_R)/2$, and $M \gg m_{L,R}$. Setting $m_R = 0$, this reduces to
\begin{eqnarray}
 \mathcal{L} &\supset&
\frac{1}{2}\bar{\Psi}_1i\gamma^{\mu}\partial_{\mu}\Psi_1 -
\frac{1}{2}M_1\bar{\Psi}_1{\Psi}_1+ \frac{1}{2}\bar{\Psi}_2i\gamma^{\mu}\partial_{\mu}\Psi_2 -
\frac{1}{2}M_2\bar{\Psi}_2{\Psi}_2
\nonumber \\
&& + i\,g_{cv}\,Z_{\mu}\,\bar{\Psi}_2\gamma_{\mu}\Psi_1
 + \frac{1}{2}g_{cv}\,Z_{\mu}\,\frac{M_1-M_2}{M_1+M_2}\,\left(
\bar{\Psi}_2\gamma^{\mu}\gamma^5\Psi_2
-\bar{\Psi}_1\gamma^{\mu}\gamma^5\Psi_1\right) +
\mathcal{O}\left(\frac{M_1-M_2}{M_1+M_2}\right)^2, \nonumber 
\end{eqnarray}
where the $M_1=M-m_+$ and $M_2=M+m_+$ are the masses of the lighter and heavier Majorana fermions, respectively. As a result of this mass splitting, all couplings between two of the same Majorana fermion and the $Z$ are suppressed by a factor of $|M_1-M_2|/(M_1+M_2) = m_+/M$, strongly limiting the rates of both self-annihilation and elastic scattering with nuclei. In contrast, interactions between the two different Majorana states, $\Psi_1\, \Psi_2$, and the $Z$ are not suppressed, potentially allowing for coannihilations to efficiently deplete their abundances in the early universe.

Throughout this study, we have assumed that there is only one relevant particle in the dark sector. In some UV complete scenarios, however, there may be exist other light particles~\cite{Freitas:2015hsa,Baek:2012se,Bhattacharya:2016qsg} which relax the constraints from direct detection experiments and from measurements of the invisible $Z$ and Higgs widths.

\section{Summary and Conclusions}
\label{conclusions}

\begin{table}[t]
\resizebox{1.0\columnwidth}{!}{
\renewcommand{\arraystretch}{1.2}
\begin{tabular}{|c|c|c|c|c|c|}
\hline 
Dark Matter & $Z$, Higgs Coupling & Direct  & Status & XENON1T &Indirect ($10^{-26}$ cm$^3/$s) 
\tabularnewline
\hline 
\hline 
Majorana Fermion & $\bar{\chi} \gamma^{\mu} \gamma^5 \chi Z_{\mu}$  & $ \sigma_{_{SD}} \sim 1$ & $m_{\chi}\sim m_Z/2$ & Yes  & $\sigma v \simeq $\, small \\
  & & & or $m_{\chi} \gsim 190$ GeV   &  Up to 440 GeV & $\sigma v \simeq 2.1-2.3$ 
\tabularnewline
 \hline
  Dirac Fermion & $\bar{\chi} \gamma^{\mu} \chi Z_{\mu}$       & $ \sigma_{_{SI}} \sim 1$ & $m_\chi  \gsim 6$ TeV  & Yes &  $\sigma v \simeq 2.1-2.3$ 
\tabularnewline
 \hline
  Dirac Fermion & $\bar{\chi} \gamma^{\mu}\gamma^5 \chi Z_{\mu}$       & $ \sigma_{_{SD}} \sim 1$ & $m_{\chi}\sim m_Z/2$ & Yes & $\sigma v \simeq $\, small \\
 & & & or $m_{\chi} \gsim 240$ GeV   &  Up to 570 GeV & $\sigma v \simeq 2.1-2.3$ 
\tabularnewline
 \hline
Complex Scalar &   $ \phi^\dagger  \overset{\leftrightarrow}{\partial_{\mu}}  \phi Z^{\mu}$, $\phi^2 Z^{\mu}Z_{\mu}$   & $ \sigma_{_{SI}} \sim 1$ & Excluded &  -- &--  
\tabularnewline
 \hline
Complex Vector &  $(X^{\dagger}_{\nu} \partial_\mu X^\nu+\rm{h.c.}) Z^{\mu}$     & $ \sigma_{_{SI}} \sim 1$ & Excluded & -- &  --
\tabularnewline
\hline 
\hline
\hline 
Majorana Fermion & $\bar{\chi} \chi H$  & $ \sigma_{_{SI}} \sim 1$ & $m_{\chi}\sim m_H/2$ & Yes &  $\sigma v \simeq $\, small
\tabularnewline
 \hline
Majorana Fermion & $\bar{\chi} \gamma^5 \chi H$  & $ \sigma_{_{SI}} \sim q^2$ & $m_{\chi}\gsim 54$ GeV & No  & $\sigma v \simeq 0.0011-3.4$
\tabularnewline
 \hline
Dirac Fermion & $\bar{\chi}  \chi H$       & $ \sigma_{_{SI}} \sim 1$ & $m_{\chi}\sim m_H/2$ & Yes & $\sigma v \simeq$\,small
\tabularnewline
 \hline
Dirac Fermion & $\bar{\chi} \gamma^5 \chi H$       & $ \sigma_{_{SI}} \sim q^2$ & $m_{\chi}\gsim 56$ GeV & No &   $\sigma v \simeq 0.0012-1.7$
\tabularnewline
 \hline
 Real Scalar &   $\phi^2 H^2$       & $ \sigma_{_{SI}} \sim 1$ & $m_{\chi}\sim m_H/2$  &  Maybe &$\sigma v \simeq 0.0012-0.019$ \\ 
 & & & or $m_{\chi}\gsim 400$ GeV   &  Up to 5 TeV & $\sigma v \simeq 2.1-2.3$
\tabularnewline
 \hline
 Complex Scalar &   $\phi^2 H^2$       & $ \sigma_{_{SI}} \sim 1$ & $m_{\chi}\sim m_H/2$  & Maybe &  $\sigma v \simeq 0.0019-0.017$ \\
 & & & or $m_{\chi}\gsim 840$ GeV &  Up to 10 TeV & $\sigma v \simeq 2.1-2.3$
\tabularnewline
 \hline
Real Vector &   $X_{\mu} X^{\mu} H^2$       & $ \sigma_{_{SI}} \sim 1$ &  $m_{\chi}\sim m_H/2$ & Maybe &  $\sigma v \simeq 0.0018-0.022$ \\
 & & & or $m_{\chi}\gsim 1160$ GeV   &  Up to 15 TeV & $\sigma v \simeq 2.1-2.3$
\tabularnewline
 \hline
Complex Vector &   $X_{\mu}^{\dagger} X^{\mu} H^2$       & $ \sigma_{_{SI}} \sim 1$ &  $m_{\chi}\sim m_H/2$ & Maybe &$\sigma v \simeq 0.0012-0.0064$ \\
 & & & or $m_{\chi}\gsim 2200$ GeV   &  Yes & $\sigma v \simeq 2.1-2.3$
\tabularnewline
 \hline
  \end{tabular}
  }
\caption{A summary of the various classes of dark matter models that we have considered in this study. For each case, we list (in the column labeled ``Status'') the range of masses (if any) that is not currently excluded experimentally. For those cases which are not already excluded, we state whether XENON1T is anticipated to be sensitive to that model. We also present the range of low-velocity annihilation cross sections that can be found in each case for masses within the currently acceptable range.}
\label{table1}
\end{table}

In this study, we have systematically considered dark matter models which annihilate through couplings to the Standard Model $Z$ or Higgs boson. Overall, we find that the vast majority of the parameter space associated with these models is ruled out by a combination of direct detection experiments (LUX, PandaX-II, etc.) and measurements at colliders of the invisible $Z$ and Higgs widths. If no detection is made, we expect experiments such as XENON1T to entirely rule out all remaining $Z$ mediated models in the near future, with the exception of fermionic dark matter heavier than $\sim$500 GeV and with primarily axial couplings. Such experiments are also expected to test all remaining Higgs mediated models, with the exception of scalar or vector dark matter with masses very near the Higgs annihilation resonance ($m_{\rm DM} \simeq m_H/2$) or fermionic dark matter with a pseudoscalar (CP violating) coupling to the Standard Model Higgs boson. Very heavy dark matter with a large Higgs portal coupling ($\lambda_{\phi H}, \lambda_{XH} \gg 1$) may also be beyond the reach of XENON1T, although LUX-ZEPLIN and other planned experiments will be able to probe such models. 

In Table~\ref{table1}, we summarize the various classes of dark matter models that we have considered in this study, listing in each case the range of masses (if any) that is not currently excluded experimentally. For those cases that are not already excluded, we list whether XENON1T is expected to have the sensitivity required to test each class of model. We also present the range of low-velocity annihilation cross sections that can be found within the currently acceptable mass range. For those models with roughly $\sigma v \gsim 3\times 10^{-27}$ cm$^3/$s (corresponding to $\sigma v \gsim 0.3$ in the units used in the Table), the Galactic Center gamma-ray excess could plausibly be generated though dark matter annihilations.

\bigskip

\textbf{Acknowledgments.} We would like to thank John Kearney for helpful conversations, and in particular for bringing to our attention an error that appeared in the first version of this paper. ME is supported by the Spanish FPU13/03111 grant of MECD and also by the European projects H2020-MSCA-RISE-2015 and H2020-MSCA-ITN-2015/674896-ELUSIVES. AB is supported by the Kavli Institute for cosmological physics at the University of Chicago through grant NSF PHY-1125897. DH is supported by the US Department of Energy under contract DE-FG02-13ER41958. Fermilab is operated by Fermi Research Alliance, LLC, under Contract No. DE-AC02-07CH11359 with the US Department of Energy. This work was carried out in part at the Aspen Center for Physics, which is supported by National Science Foundation grant PHY-1066293.

%%%%%%%%%%%%%%%%%%%%%%%%%%%%%%%%%%%%%%%%%%%%%%%%%%%%%%
\bibliography{higgsz.bib}
\bibliographystyle{JHEP}
%%%%%%%%%%%%%%%%%%%%%%%%%%%%%%%%%%%%%%%%%%%%%%%%%%%%%%

\end{document}